\documentclass[aps,prl,amsmath,amssymb,reprint]{revtex4-2}
\usepackage{lmodern}
\usepackage{graphicx}
\usepackage{dcolumn}
\usepackage{bm}
\usepackage[utf8]{inputenc}
\usepackage[T1]{fontenc}
\usepackage{mathptmx}
\usepackage{etoolbox}

\usepackage[english]{babel}

\begin{document}
\title{Intrinsic Dimensionality of Molecular Properties}
\author{Ali Banjafar}
\affiliation{Institut f\"ur Chemie, Universit\"at Kassel, Heinrich-Plett-Stra{\ss}e~40, 34132~Kassel,~Germany}
\author{Guido Falk von Rudorff}%
\email{vonrudorff@uni-kassel.de}
\affiliation{Institut f\"ur Chemie, Universit\"at Kassel, Heinrich-Plett-Stra{\ss}e~40, 34132~Kassel,~Germany}
\affiliation{Center for Interdisciplinary Nanostructure Science and Technology (CINSaT), Heinrich-Plett-Stra{\ss}e 40, 34132 Kassel}

\date{\today}
\begin{abstract}
Chemical space which encompasses all stable compounds is unfathomably large and its dimension scales linearly with the number of atoms considered. The success of machine learning methods suggests that many physical quantities exhibit substantial redundancy in that space, lowering their effective dimensionality. A low dimensionality is favorable for machine learning applications, as it reduces the required number of data points. It is unknown however, how far the dimensionality of physical properties can be reduced, how this depends on the exact physical property considered, and how accepting a model error can help further reducing the dimensionality. We show that accepting a modest, nearly negligible error leads to a drastic reduction in independent degrees of freedom. This applies to several properties such as the total energy and frontier orbital energies for a wide range of neutral molecules with up to 20 atoms. We provide a method to quantify an upper bound for the intrinsic dimensionality given a desired accuracy threshold by inclusion of all continuous variables in the molecular Hamiltonian including the nuclear charges. We find the intrinsic dimensionality to be remarkably stable across molecules, i.e. it is a property of the underlying physical quantity and the number of atoms rather than a property of an individual molecular configuration and therefore highly transferable between molecules. The results suggest that the feature space of state-of-the-art molecular representations can be compressed further, leaving room for more data efficient and transferable models.
\end{abstract}

\maketitle

\section{Introduction}

When characterizing chemical space, one of its fundamental properties is how many independent dimensions it has. That is, to quantify how many independent variables are required to describe how a property changes between molecules and molecular configurations. Formally, chemical space has $4N$ continuous dimensions where $N$ is the number of atoms. This can be seen from the components of the molecular Hamiltonian $\hat{H}(\mathbf{R}_I, Z_I)$ which depends on the nuclear charges and positions of each atom. Translational and rotational symmetry reduces the number of dimensions by 5 or 6 for linear and non-linear molecules, respectively.

The wealth of research and evidence from machine learning that many molecular properties can be modeled and predicted based on exploiting similarities in chemical space\cite{keith2021combining,Lu2024,VonLilienfeld2020} is indicative of chemical space being highly redundant and thus compressible for many properties of interest. Since machine learning approaches are data-driven and thus require large number of data points\cite{Smith2017a,Ganscha2025}, which in turn are costly to generate, it is particularly interesting to learn how far chemical space can be compressed without or minimal loss of predictive power.
This is particularly relevant for machine learning applications, since learning theory suggests that the number of data points required to approximate an arbitrary but well-behaved function to some given accuracy (also) depends on the dimensionality of said function\cite{Madych1992}. The dimensionality of a mathemtical pure function would be represented as the number of (possibly redundant) arguments that it takes. Therefore, machine learning representations can be seen as a parameter transform from the original Cartesian coordinates and nuclear charges of the molecular Hamiltonian into another vector space that is more amenable to interpolation between the values of a given property.

Those representations can either be generated implicitly during training of a neural network\cite{Krenn2022} or they can be designed based on physical insight\cite{Musil2021,David2020}. 
Either way, these representations contain typically on the order of a few hundred to two thousand entries or dimensions\cite{Briling2024,Karandashev2022,Faber2018,Khan2023}. Given typically molecular atom counts $N$, this is substantially more than four $N$. This over-completeness can be beneficial since it allows to learn several different properties using the same representation even though grouping in similarity would be different depending on property. However, representations benefit from being as short and compact as possible. For example, in kernel-ridge regression or other kernel methods, distances between representation vectors in training and test data have to be evaluated which scale linearly with the number of data points. Since kernel methods are not scale invariant, each additional feature formally introduces another hyperparameter by which it can be scaled. Consequentially, the search space of the hyperparameter optimization in turn scales with the number of features. In practice, this is addressed by setting a parameter during development of the representation and not scaling the features for the individual application except for categorial regression, i.e. one-hot-encoding. Longer representations severely impact the time to result.

Moreover, applications of confidential computing or multi-party computing, which may allow predictions on confidential data, benefit from compact representations\cite{Weinreich2023a}. With recent efforts into short yet transferable machine learning representations\cite{Briling2024,Khan2023}, it is an important question how far the number of independent degrees of freedom can be reduced when describing chemical space. Naturally, that number must depend on the property under consideration. For example, the net charge is independent of the configuration, but depends on the nuclear charges of all atoms while the surface area depends on the positions but not the nuclear charges. In thus far, the dimensionality is less of a property of chemical space as a whole, but rather of a property. Consequently, it is important to be able to quantify the limit of the minimal number of degrees of freedom required to describe a physical property based on the data or the variance of that property alone. This would allow to build more data-efficient models tailored to learn specific properties, if the limit of that number of degrees of freedom is known and might explain why some properties are easier to learn than others.

The dimensionality of a property in chemical space has a direct physical connection. If a property is nearsighted, as it is often invoked for example for total energies, then that would indicate that there should be a finite number of degrees of freedom that can impact that property in order to preserve locality. If a property was global instead, then it would need to scale with the overall system size. In machine learning terminology, introducing a cut-off threshold beyond which interactions are not taken into consideration directly implies locality, which in turn implies reduction of dimensionality. Identifying an upper bound of said dimensionality therefore is identical to identification of a lower bound of compressibility or dimensionality reduction for machine learning applications.

In recent decades, advancements in computational chemistry and machine learning have significantly heightened interest in intrinsic dimensionality and dimension reduction concepts \cite{keith2021combining,  jia2021feature, vandermaaten2009dimensionality}. Intrinsic dimension refers to the minimum number of variables that minimize information loss in a data set or physical properties \cite{camastra2016intrinsic}. With an intrinsic dimensionality estimate, often a dimensionality reduction scheme can be implemented\cite{orlov2025exploring}. In machine learning, this connects to active learning, feature selection, dimensionality reduction \cite{tamilselvi2017review}, for the dimensionality estimation of point clouds and for cluster identification \cite{Williams2018}.

Typically dimensionality reduction implies a loss of precision\cite{Laikov2011}. This also applies to quantum chemistry, where sparsifying interactions has been a successful strategy\cite{Schleder2019} in method development: only rarely this can be done without loss of accuracy as it has been done e.g. in equating knowing the wavefunction (which has $3N_\textrm{e}$ dimensions) with knowing the electron density (which has 3 dimensions, independent of the number of electrons  $N_\textrm{e}$) by virtue of the Hohenberg-Kohn theorem. 

Literature distinguishes global intrinsic dimension (ID) and local ID. The former gives the number of degrees of freedom to approximately describe the global shape of, for example, a point cloud, while the latter describes the region around a certain point\cite{camastra2016intrinsic}. The global ID is the same everywhere, while the local ID can differ. Since we typically only characterize small regions of chemical space with a certain application in mind without ever building a random sample of all possible molecules, the local ID is most suitable for characterizing chemical space. 

Point-cloud ID estimators rely on the geometry of the point cloud and the relationships between inter-point distances. For example, they often examine the distribution of distances to nearest neighbors\cite{Ceruti2014, Albergante2019}. These methods are typically sensitive to the density and distribution of points \cite{Facco2017, Denti2022} -- issues that become particularly problematic in high-dimensional spaces \cite{Gomtsyan2019}. To estimate the ID of a physical property using such methods, one would need to generate sample points on a level set (constant value surface) in a $4N$-dimensional space, estimate the dimensionality of that surface (akin to the null space) and subtract it from the full number of dimensions of the embedding space. This however poses significant challenges in terms of both the required number of points and maintaining uniform point density across the space.

Among ID estimators, Principal Component Analysis (PCA)-based methods have received significant attention \cite{Bouveyron2011, Rozza2012, Erba2019}. In the field of chemistry, PCA is commonly used as a tool to identify and quantify the most relevant variables in molecular systems. However, in practice, the application of PCA is often indirect by first defining a representation and then identifying and counting key components by PCA\cite{Grunwald2015,Shukur2011}, which naturally heavily depends on the choice of the representation\cite{Le2018}. Typically, the result of a PCA is a global ID, rather than a local one and moreover requires the feature space to be linearizable which it almost never is. A notable exception for nonlinear PCA would be the kernelized variant, which still yields a global ID.

The local ID is conceptually related to the tangent space approach, where a flat surface defined by the first derivatives of the target function is constructed at a specific point on the surface. The tangent vectors at that point represent the local directions of change relative to the reference point. Therefore, it is a local ID picture, since changing the position alters the tangent space and potentially its number of dimensions\cite{Yao2022}. The tangent space however is restricted to considering a single point in chemical space, and does not directly allow for investigation of the accuracy-ID tradeoff, as it reproduces the formal ID for infinite accuracy.

Our method does not rely on a locally flat tangent plane at a single reference point. Instead, we approximate a region within the thermally accessible space using a Taylor expansion of the property surface. This approximation captures both the local slope and curvature through the gradient and Hessian, respectively. Unlike tangent vectors, the eigenvectors and eigenvalues of the Hessian matrix describe the principal directions and magnitudes of the surface curvature. As a result, our method allows for accurate modeling of the property surface over a broader region without requiring movement of the reference point or repeated recalculations and allows for simple detection of (approximate) symmetries.

\begin{figure*}
    \centering
    \includegraphics[width=\textwidth]{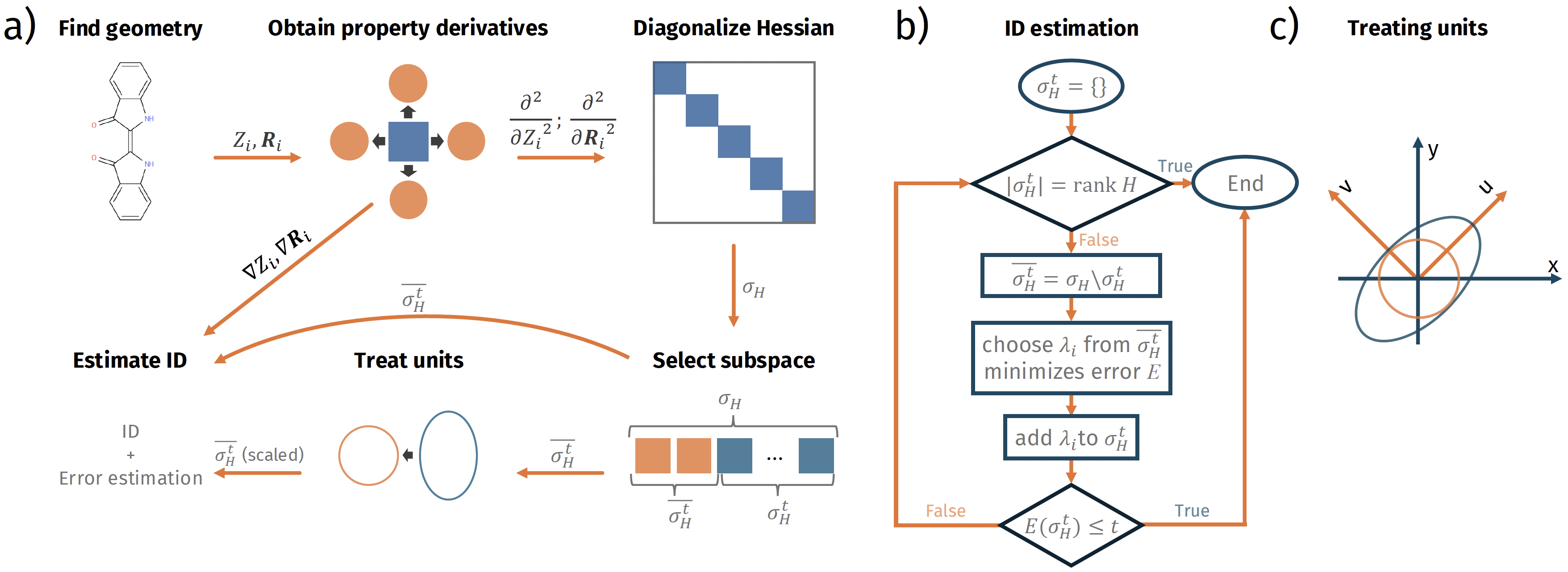} 
    \caption{Illustration of the estimation of the intrinsic dimension (ID) of a physical property. a) Workflow: at a given molecular geometry we evaluate property derivatives, building the gradient $\nabla\cdot$ and the Hessian matrix $H$. Selecting a subspace of the diagonalized Hessian enables to assess the ID and its corresponding approximation error. 
    b) Flow chart of the joint subspace selection and error estimation process. 
    c) Visual representation how the difference of units between spatial and charge degrees of freedom can be used to reduce dimensionality through scaling eigenvectors.
    }
    \label{fig:schematic}
\end{figure*}

\section{Methods}

In this work, the goal is to determine the local intrinsic dimension of physical properties such as the total energy, HOMO-LUMO gap and HOMO orbital energy for various molecules at a given geometric configuration, not necessarily a local minimum. \textit{Local} means that we consider the intrinsic dimension to be dependent on both the particular molecular configuration and the property. Even though the number of dimensions depends on accuracy requirements, our approach is general as it yields the minimal number of dimensions for a given error bound.

We first build a Taylor expansion around the molecular configuration and then analyze the resulting multivariate polynomial. The polynomial depends on all nuclear charges $Z_I$ and nuclear positions $\mathbf{R}_I$, so it has $4N$ variables for $N$ atoms.

\subsection{Ab initio calculation and Taylor expansion}
While the general form of a multivariate Taylor expansion of a property $p({\boldsymbol {x}})$ with all degrees of freedom of $N$ atoms merged into vector ${\boldsymbol {x}}\equiv\mathbf{R_1}\oplus\dots\oplus  \mathbf{R_N}\oplus Z_1\oplus \dots\oplus Z_N$ is given by:
\begin{align}
    p({\boldsymbol {x}})\simeq \sum _{|\alpha |\leq k}
    \frac {\partial ^{|\alpha |}p(\mathbf{a})}{\partial x_{1}^{\alpha _{1}}\cdots \partial x_{n}^{\alpha _{n}}}
    \frac{(\mathbf {x}-\mathbf {a})^\alpha}{\alpha!}.\label{eq:Taylor}
\end{align}
with the multi-index $\alpha$ and the reference molecule $\mathbf{a}$. In this work, we use a second-order expansion (i.e. $k=2$). Unification of the nuclear charge degrees of freedom and the spatial degrees of freedom is inspired by the quantum alchemy approaches\cite{von_Rudorff_2020,Rudorff2021a,Lilienfeld2009}, including alchemical normal modes\cite{Fias2018}, which in turn continue in the spirit of the four-dimensional electron density concept\cite{Wilson1962} or conceptual DFT\cite{Ayers2009,Balawender2013}.

Similar derivatives have been considered with Hartree-Fock\cite{Lesiuk2012,Rudorff2021a,TamayoMendoza2018}, DFT\cite{Lilienfeld2006,Lesiuk2012,Balawender2013,Griego2018,Kasim_2022,Munoz2020,MirandaQuintana2017} and CCSD\cite{von_Rudorff_2020,Abbott2021}, so we expect our approach to be applicable for many levels of theory. In this work, we consider Restricted Kohn-Sham Density Functional Theory (RKS-DFT) with the PBE\cite{Perdew1996} exchange-correlation functional and Restricted Hartree-Fock, as implemented in PySCF\cite{PYSCF}. Since all the molecules considered in this study are neutral closed-shell systems, RKS is appropriate. 
We use the uncontracted cc-pVQZ basis set to reduce artifacts from basis functions being developed for integer nuclear charges, which is known to affect response functions since the Hellmann-Feynman theorem is not satisfied if the basis functions are not sufficiently flexible in the direction of changed of nuclar charges\cite{Domenichini2020}. 

\subsection{Analytical and Finite differences derivatives}
Ideally, all derivatives would be calculated analytically. While spatial derivatives are widely implemented in quantum chemistry codes via Coupled-Perturbed (CP) approaches, only few implementations are available for alchemical derivatives either following CP approaches\cite{Lesiuk2012,Domenichini2022},  automatic differentiation \cite{Kasim_2022,TamayoMendoza2018,Zhang2022} or  arbitrary precision operations\cite{Rudorff2021a}. Implementations of analytical derivatives are currently limited to derivatives of the total energy as the property of interest and restricted Hartree-Fock with the main exception being the first order derivative of the orbital eigenvalues which are a by-product of the CP method. Higher orders have been described both for spatial\cite{Osamura1986} and alchemical\cite{Lesiuk2012} derivatives, but no implementation is available, so numerical differentiation is used instead, which also allows to accept some roughness of e.g. a DFT property surface for high order derivatives\cite{Zhou2023}.

In this work, we use analytical derivatives where possible (i.e. for energies) and numerical derivatives on all other cases as implemented in our unifying open-source python package \texttt{nablachem.anygrad}\cite{Banjafar2025}.

\subsection{Estimating the local intrinsic dimension}

The primary goal of this work is to calculate the local intrinsic dimension of various properties across a wide range of molecules. To achieve this, we leverage the fact that molecular properties are  smooth functions of their spatial and chemical coordinates within chemical space. This allows us to examine the property surface in the vicinity of a fixed molecular configuration. By analyzing the shape and curvature of the property surface, we can identify the key coordinates that play a significant role in defining these properties.

We define the local intrinsic dimension of a molecular property $p(\mathbf{x})$ (with $\mathbf{x}$ defined as in eqn~\ref{eq:Taylor}) in the domain $\Omega$ around a molcular configuration as the minimal set of orthogonal vectors $\sigma_H^t$ used either as gradient or as Hessian eigenvector of approximant $\tilde{p}(\mathbf{x}|\sigma_H^t)$ s.t. $\left\langle \left(p(\mathbf{x})-\tilde{p}(\mathbf{x})\right)^2\right\rangle_\Omega < t$.

As input, this requires only the gradient and the Hessian of a function. The shape of the property surface is primarily determined by the eigenvalues and eigenvectors of the Hessian matrix. The threshold $t$ allows to investigate local intrinsic dimension based on different accuracy requirements.

The full second order approximation $\tilde{Q}$ of a property $Q$ is given by 
\begin{multline}
\left.
\tilde{Q}(\boldsymbol{\Delta x}) \equiv Q_0 + \sum_{i} \ \boldsymbol{\Delta x_i} \frac{\partial Q(\boldsymbol{x})}{\partial x_i} \right|_{\boldsymbol{x} = \boldsymbol{a}} \\
\left.+ \frac{1}{2}\sum_{i} \sum_{j}  \boldsymbol{\Delta x_i} \boldsymbol{\Delta x_j} \frac{\partial^2 Q(\boldsymbol{x})}{\partial  x_i \partial x_j} \right|_{\boldsymbol{x} = \boldsymbol{a}}
\label{eq:taylor_approximation}
\end{multline}

with \(\boldsymbol{\Delta x}\equiv \boldsymbol{x} - \boldsymbol{a}\), where \(\boldsymbol{a}\) is the vector encoding nuclear charges and positions of a given molecular configuration, and \(\boldsymbol{x}\) represents the coordinates of an arbitrary point close-by.  To identify the most relevant number of eigenvectors (\(k\)) for a given property surface, we select a subset of the unordered eigenvalues $\lambda_i$ and eigenvectors $\mathbf{y}_i$ of the Hessian matrix. 

\begin{multline}
\left.
\tilde{Q'}(\boldsymbol{\Delta x}) \equiv Q_0 + \sum_{i} \ \boldsymbol{\Delta x_i} \frac{\partial Q(\boldsymbol{x})}{\partial x_i} \right|_{\boldsymbol{x} = \boldsymbol{a}} 
+ \sum_{i}^k  \lambda_i (\boldsymbol{\Delta x}^\textrm{T}  \boldsymbol{y_{i}} )^2
\label{eq:taylor_re-approximation}
\end{multline}
Note that the selection of eigenvectors $\mathbf{y}_i$ is done by minimizing the approximation error and not by choosing the largest eigenvalues $\lambda_i$ alone. This way we include the whole relevant approximation neighborhood $\Omega$ in the model objective and can thus quantify the actual approximation error.

While the full search space of subsets scales exponentially with the number of available eigenvectors to choose from, we employ a greedy algorithm (see below) to provide a linear-scaling estimate of the optimal solution. Comparison to random sampling of eigenvector sets of given size $k$ for small molecules confirmed that the global optimum is close to the greedy search results in all our tests. The iterations (see also Fig.~\ref{fig:schematic}) begin with \(k = 0\), which corresponds to a linear approximation. At this stage, only the reference value \(Q_{0}\) and the gradient \(\nabla Q\) are considered in the Taylor expansion. With each iteration, \(k\) is incremented by one. However, the pair of eigenvalue and eigenvector included at each step must be those that have the strongest influence on the property surface. In terms of the error estimation process, the newly added eigenvalue-eigenvector pair minimizes the difference between the second-order Taylor approximation of the property with the full Hessian matrix and the approximation considering only \(k\) eigenvalues and eigenvectors if averaged over the local neighborhood volume $\Omega$.

During each iteration, we examine whether the gradient vector of the property is already described by the those eigenvectors of the Hessian matrix that have been selected for inclusion. Specifically, at iteration $k$, we project the normalized gradient onto all $k$ selected eigenvectors and check whether the remaining gradient is a non-zero vector. 
If it is non-zero, it represents an additional degree of freedom, meaning that the estimated intrinsic dimension is $k+1$.

\subsection{Error Estimation}

In the error estimation, the ground truth is compared to the model \(\tilde{Q}'\) over a finite domain \(\Omega\) which is the neighborhood of the molecule in chemical space. We consider the thermally accessible region as a physically meaningful definition of locality and thus model the boundary of $\Omega$ by requiring the energy difference being less than $5k_BT$. 

Conceptually, this comparison should ideally be made with respect to the underlying ab initio model (e.g., DFT). However, such an approach would require numerical integration, which is computationally demanding and unnecessarily expensive. Since this work focuses on the local intrinsic dimension, using the full second-order model \(\tilde{Q}\) as the comparison is preferred. This choice is justified because:  a) \(\tilde{Q}\) is extremely close to the ab initio model, and  b) \(\tilde{Q}\) allows for analytical integration.  As shown in the supporting material (Fig. S1), the difference between the ground truth \(Q\) and the second-order approximation \(\tilde{Q}\) at the boundaries of the integration domain \(\Omega\) is negligible. We choose the root-mean-square error (RMSE) as error metric because it can be evaluated analytically:
\begin{align}
\textrm{RMSE}\equiv\sqrt{\langle\tilde Q'-\tilde Q\rangle^2_\Omega} = \sqrt{V_\Omega^{-1}\int_\Omega (\tilde Q'(\mathbf{x})-\tilde Q(\mathbf{x}))^2\,d\mathbf{x}} 
\end{align}

\subsection{Treating units and rotational symmetries}
The molecular Hamiltonian includes parameters with two different units: spatial degrees of freedom and nuclear charges. To allow identification of symmetries in the eigenvector space, we introduce a conversion factor which removes the spurious degree of freedom that originates from the units only and has no physical meaning. This conversion factor $s\neq 0$ is applied to all nuclear charge units equally and is found again by minimizing the resulting number of dimensions.

Finally, we remove rotational symmetries in the diagonalized frame (see Fig.~\ref{fig:schematic}). In a centrosymmetric function given by $x_1^2+x_2^2$, we only have one relevant degree of freedom, the radius $r^2$. Such a \textit{continuous} symmetry is only possible if the coefficients for monomials $x_i^2$ are identical: otherwise, the function would become an ellipsis and we would need to know both components $x_i$ to determine the function value, not only the radius $r^2$. This is equivalent to having no mixed terms—terms that arise from the multiplication of different degrees of freedom. In our approach, rotational symmetry is detected when the diagonal form contains degenerate eigenvalues.

Combining the individual steps in the sequence of this section (cf. Fig.~\ref{fig:schematic}), we obtain our estimate of the local intrinsic dimensionality with an associated RMSE over the neighborhood $\Omega$.

\section{Results}
\begin{figure*}
    \centering
    \includegraphics[scale=0.4]{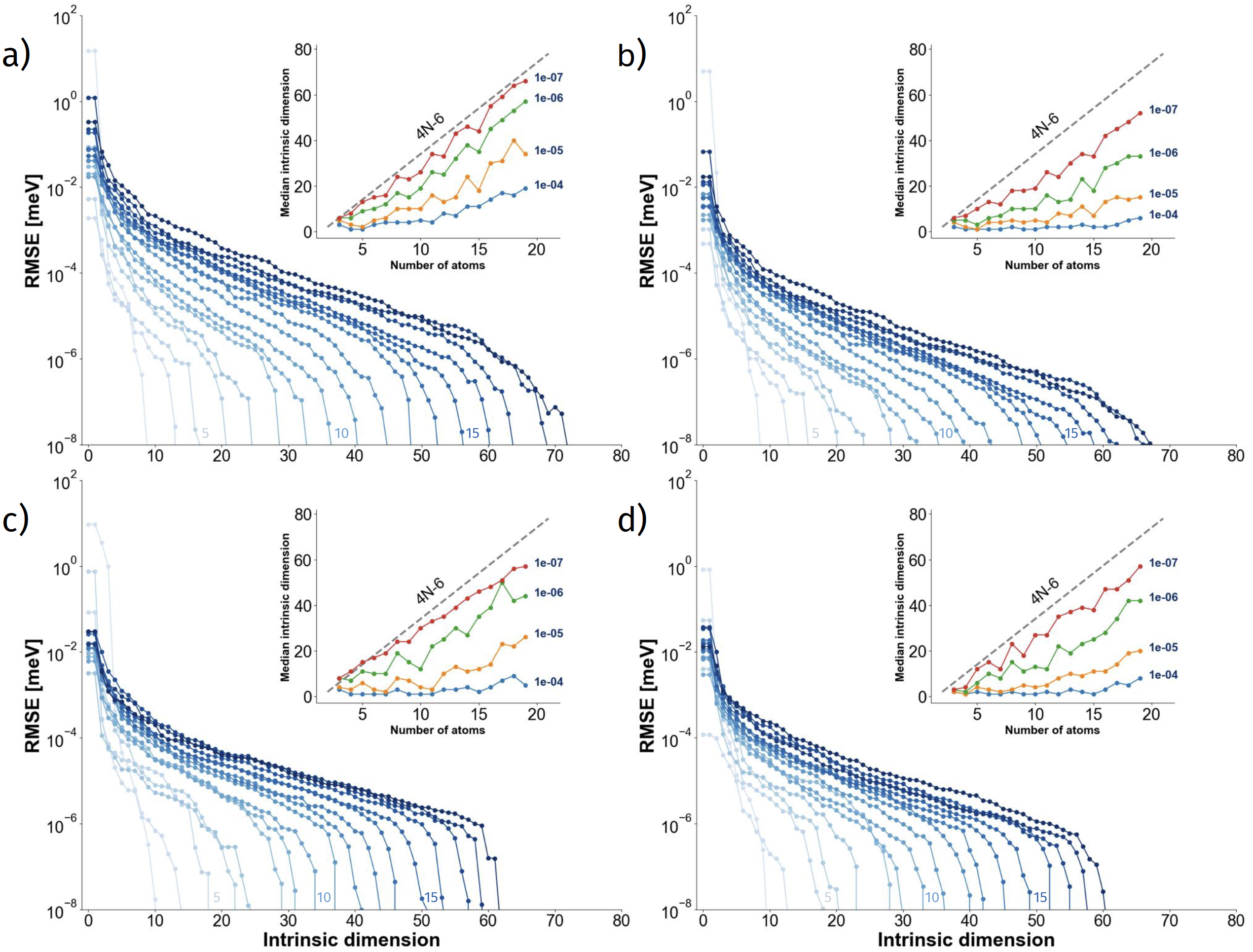} 
    \caption{Intrinsic dimensions (ID) and the corresponding approximation error for a) total energy, b) total energy per number of atoms, c) HOMO-LUMO gap, and d) HOMO orbital energy. 
    Each line shows the median ID value for molecules of the same number of atoms (color darkens with molecule size, every fifth entry is annotated with the number of atoms). 
    tual estimated ID points, which are connected to highlight the trend more clearly. 
    Insets: Median ID with number of atoms for different accuracy levels (in meV) Upper bound thereof given by vibrational degrees of freedom as dashed line.}
    \label{fig:idscan}
\end{figure*}

We applied our method across a random subset of 370 neutral molecules (those where the DFT geometry optimization converged out of 1,000 initially sampled molecular graphs) with $< 20$ atoms from ChEMBL\cite{Mendez2019} for the total energy (an extensive property) as well as the HOMO-LUMO gap and HOMO orbital energy (both intensive properties). 

The results of the estimation for the intrinsic dimensionality of the total energy of the molecules is shown in Fig.~\ref{fig:idscan}a. To facilitate the comparison of the total energy to intensive properties, we also consider the total energy normalized by the number of atoms, making it intensive-like (see Fig.~\ref{fig:idscan}b).

The Pareto front of the lowest root-mean-square error (RMSE) of all considered physical quantities attainable given a certain intrinsic dimension follows a remarkably similar shape, as shown in Fig.~\ref{fig:idscan}. The overall shape can be understood as three different regimes: the initial steep improvement of the error as the first few (fewer than 10) dimensions are added, the long and slow decay plateau and the steep drop for the final few (less than 10) dimensions. We will understand the three domains as \textit{separable}, \textit{coupled} and \textit{redundant} dimensions, respectively.

The \textit{separable} dimensions form the first region which features a initial steep decrease in error as the number of intrinsic dimensions increases. This regime reflects the significant impact of the primary degrees of freedom on the calculated property, where the most influential eigenvalues and eigenvectors can be well-separated from the rest. This is akin to principal components which describe the dominating directions in a vector space PCA, and captures only those components which are linear or quadratic in the cartesian and nuclear charge dimensions. Since the scaling factor removes the ambiguity of the units between nuclear charges and coordinates and the removal of degenerate eigenvectors due to their nature of rotational degrees of freedom, the number of independent dimensions can be substantially lower than the (complete) set of all alchemical normal modes\cite{Fias2018}, further illustrating redundancy in chemical space.

The \textit{coupled} domain characterized by a relatively flat but never stagnant decay of the error with additional dimensions indicates that a complex non-separable interaction of the many degrees of freedom is slowly and inefficiently expanded in second order terms. The monotonous decay of the median error with additional dimensions points towards this expansion being well behaved. However, this expansion does not include degrees of freedom which can be expressed as single linear combination of the cartesian nuclear coordinates and their charges, since those degrees of freedom would already be covered in the domain of the separable dimensions. The overall decay of that plot is in line with common exponential eigenvalue decay rates found in physics applications\cite{Fyodorov2018,Dressler2020,DeHoop2007} and machine learning applications\cite{Shawe-Taylor2005} for general or random functions.

Finally, the \textit{redundant} dimensions are reached. Ideally, the last six (or five for linear systems) degrees of freedom are zero due to the translational and rotational symmetries of the molecules. However, due to the finite precision of the underlying \textit{ab initio} calculations, these last six values are only close to zero. Even if the Hessian matrix is obtained analytically via coupled-perturbed methods, the self-consistent iterations are only continued until a finite convergence threshold. Any deviation of the RMSE from zero for the last five dimensions is solely a consequence of that effect.

\begin{figure}
    \centering
    \includegraphics[scale=0.25]{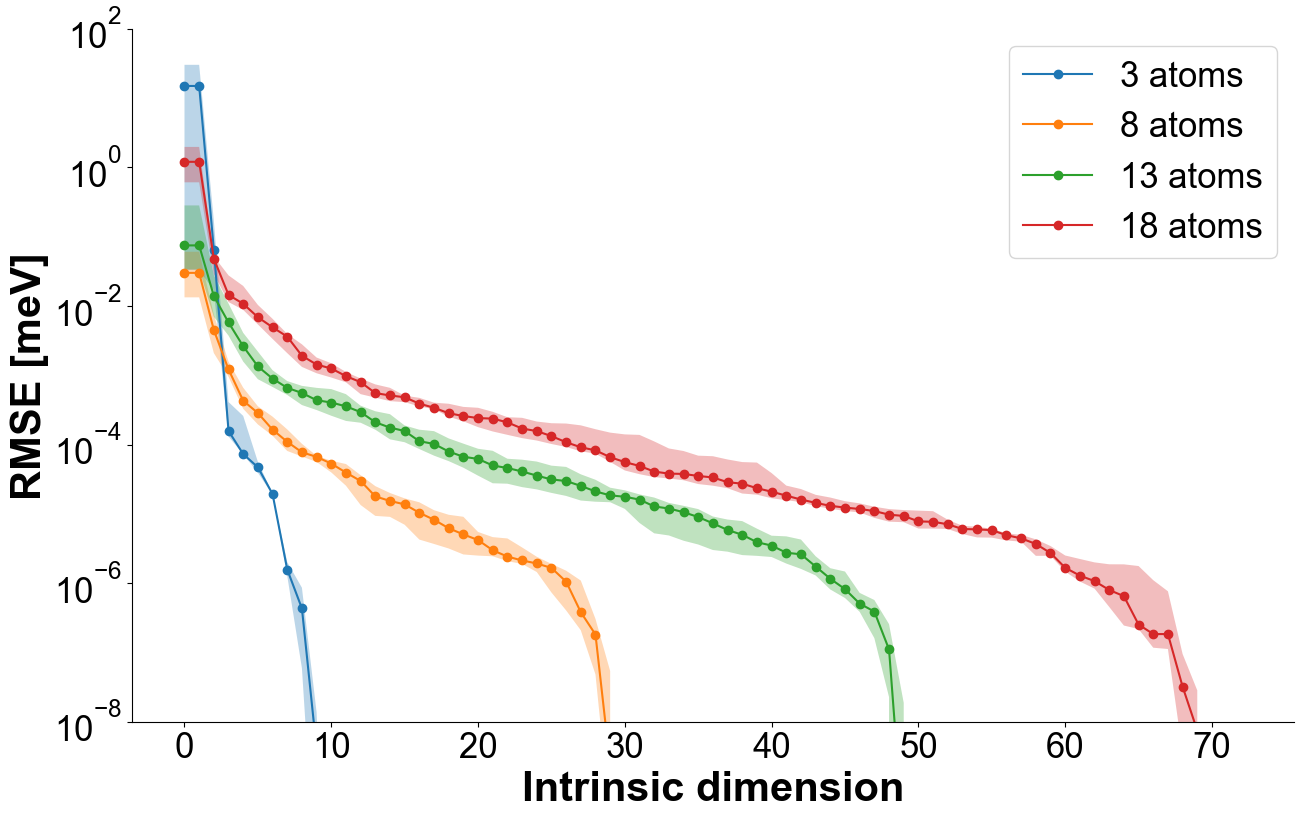} 
    \caption{Variability of the relationship between number of intrinsic dimensions and the resulting accuracy over several molecules of the same number of atoms. Dots indicate the median, shaded area the 33$^\textrm{rd}$ until 67$^\textrm{th}$ percentile. All data for the total energy. For legibility, only a few atom numbers are shown.}
    \label{fig:percentile}
\end{figure}

Since Fig.~\ref{fig:idscan} shows the median RMSE for the total energy over all molecules of a given number of atoms, it is interesting to note how these results differ between molecules of the same size. Fig.~\ref{fig:percentile} exemplifies this for the total energy and different number of atoms. It becomes evident that the variance of the RMSE, a forgiven number of intrinsic dimensions, is very low across molecules of same size. The largest variance is to be found for the separable domain, where the individual molecules exhibit different offsets. Those are the consequences of the strength of the curvature for those systems. Note that this variance also manifests itself in the dependency of that offset on molecular size. Additional random sampling of molecules might reveal a trend in this offset. In the region of coupled dimensions, we typically see a low variance or the approximation error between molecules of same size, which is in line with the interpretation of that region describing nonlinear and non-quadratic interactions between the spatial and charge degrees of freedom. It is only towards the domain of redundant dimensions that the variance widens. This is in line with this region being dominated by numerical noise of the Hessian matrix which will have a strong dependence on the particular molecule. The large variance in the zero order term, i.e. the total energy for a static configuration, is not visible in Fig.~\ref{fig:idscan}, since adding a constant offset to the property value is a change of dimension zero.

The results in Fig.~\ref{fig:percentile} imply that the choice of specific molecules for calculating the ID does not significantly influence the results except for the separable region, making the results likely transferable for the chemical space of small neutral and stable molecules. While the data in Fig.~\ref{fig:percentile} is shown for the total energy, the trends of the results are similar for the HOMO-LUMO gap and the HOMO alone. This is particularly remarkable since charged compounds or non-covalently interacting systems\cite{Stohr2019,Fedik2022} may exhibit substantially longer-ranged interactions, the relative extend of the three regimes may prove different in those systems.

\subsection{Intrinsic dimension of total energy}

When normalizing the total energy into the total energy per atom, the property becomes almost intensive, as the scaling with the size of the system is approximately removed. 
In Fig.~\ref{fig:idscan}, the distinction between the extensive and intensive perspective manifests as a downward shift in ID median values. The relevance of this difference however becomes more clear when considering the scaling behavior of the ID with number of atoms (the insets in Fig.~\ref{fig:idscan}). Formally, one would expect the ID to scale with $4N-6$ to reflect the total number of degrees of freedom. This is mostly the case for extremely high accuracy requirements of $10^{-7}$ meV, which are common convergence thresholds for \textit{ab initio} calculations. While the overall linear increase of ID with number of atoms persists for different precisions from $10^{-4}$ to $10^{-7}$ meV, the slope decreases, i.e. larger molecules can be treated with only a reduced number of additional dimensions. 

For the energy per atom in Fig.~\ref{fig:idscan}b, an interesting feature appears: for intermediate accuracy requirements, the ID remains constant after a minimal size of about 15 atoms, as indicated by the plateau of the ID for 15-20 atoms e.g. for RMSE thresholds of $10^{-5}$ or $10^{-6}$. This  behavior is in line with the expected locality of that property, sometimes called the \textit{nearsightedness of matter}\cite{Prodan2005}. In this work, we can quantify an upper bound for the number of atoms where this happens for the energy of small neutral molecules: the aforementioned 15-20 atoms. Analyzing the individual degrees of freedom to better understand the nature of the finite degrees of freedom is beyond the scope of this work and requires a more extensive sampling of chemical space. Note that the dimensions we find include collective degrees of freedom, so our results are what one would expect if locality is present but themselves do not imply locality. For example, non-covalent interactions can be described by few collective degrees of freedom, each of which are coupling a large domain of the system\cite{DiStasio2012}. This way, even long-range effects may be of low intrinsic dimensionality.

Naturally, if no approximation of the underlying property is allowed, all formal $4N-6$ degrees of freedom become required.

\subsection{Intrinsic dimension of HOMO-LUMO gap and orbital energies}
The nature of the HOMO-LUMO gap and the total energy is fundamentally different. While the energy is both extensive and often local for neutral molecules, the potentially delocalized nature of molecular orbitals renders the HOMO-LUMO gap less likely to be compressible in dimensionality reduction. This has rendered direct learning of either quantities a substantial challenge which is either addressed with tailored representations\cite{Karandashev2022} or exploited by using their sensitivity in modeling and representations\cite{Welborn2018,Zulueta2022,Briling2024}. 

Mathematically speaking, the HOMO-LUMO gap is bounded from below by zero unlike the molecular energy which should affect the results for molecules with a narrow gap. In direct comparison in Fig.~\ref{fig:idscan}, we find that much fewer intrinsic dimensions are needed to describe the gap compared to the total energy. This suggests that the gap should be easier to describe since only lower complexity surface is to be modeled. At the same time the decay especially on the coupled regime is much flatter highlighting substantially stronger coupling between the degrees of freedom such that the marginal accuracy gain by adding additional dimensions is diminished compared to the extensive energy. This can be understood as the consequence of having many coupled degrees of freedom with more similar eigenvalues: the flatter the decay in the coupled regime, the more likely that two formally independent directions can be folded into one by symmetry, so a perfectly flat behavior is impossible. However, since the approximation error (and not the eigenvalues alone) decay only slowly, this indicates that the many degrees of freedom are highly coupled and form a non-symmetric balance in describing the local environment akin to an alternating sum. This might contribute to the HOMO-LUMO gap being a comparably hard machine learning application. We observe the same effect for all accuracy levels (see inset in Fig.~\ref{fig:idscan}c), so reducing accuracy does not qualitatively affect learning complexity.

For gap (Fig.~\ref{fig:idscan}c) and orbital energy (Fig.~\ref{fig:idscan}d), trends and values of the IDs behave similarly, implying that the question of orbital occupancy minimally affects the ID value. This is in line with the conceptual model of the underlying physical problem: molecular orbitals can be rotated and within this rotation preserve the eigenvalue spectrum but not shape and localization. Thus, they require delocalized and global support to exhibit that property. As such, frontier orbitals serve as global molecular fingerprint since they are particularly sensitive to small-charge redistributions. Here we find this to be the case also including alchemical degrees of freedom, not only spatial ones.

\section{Conclusion}
In this work we present a method to quantify the trade-off between accuracy requirements and the intrinsic dimension of a given property in chemical space. This serves as upper bound for the minimal length of machine learning representations describing said properties for a variety of molecules and can be substantially below the formal dependency of $4N$. While there are recent efforts in shortening machine learning representations for efficiency\cite{Khan2023} and scalability\cite{Huang2023}, our results suggest that even the current representations are overcomplete. While redundancy might render representations more general as they offer different features that may be relevant to learn several properties using the same representation, our work suggests that using more compact representations for individual properties should be generally possible. This is desirable because this reduces not only the total number of data points needed in order to train a model of certain accuracy but also is expected\cite{Stone1982} to improve the data efficiency (i.e. the marginal prediction improvement from a single additional data point).

Our work further suggests that relaxed accuracy requirements may be one way to reduce the intrinsic dimensionality for the otherwise unchanged chemical space. Since reduced dimensionality impacts the learning behavior of models\cite{Stone1982}, this in turn suggests that reducing accuracy requirements should improve the data efficiency of machine learning models if considered at the design stage thereof. In this context, it is important to note that a strong reduction of the intrinsic dimensionality can be achieved already at relaxed accuracy thresholds which are much lower than the requirements of most computational chemistry applications.

We find that frontier orbital energies and their difference couple degrees of freedom much more strongly than total energies, which is in line with the expectation of orbital energies to depend on the overall molecular structure. It is remarkable that the median estimate of the approximation error depends largely on the underlying property and the molecular size, but not on the molecule and its topology. This indicates that different physical properties exhibit fundamentally different complexity which only emerges in the data-driven perspective. This behavior is surprisingly transferable between molecules, which makes it interesting for applications\cite{Huang2023}.

The main limitation is that we consider the local environment around each molecular minimum energy configuration only. Non-equilibrium configurations could have a higher coupling between the degrees of freedom, since in that case, the total energy can be kept constant by increasing the energy along some dimensions and decrease it along others, which is impossible for equilibrium configurations.

\section{Data Availability}
All data is available online\cite{Banjafar2025} and the code is available as part of our \cite{Banjafar2025a} Python package.

\bibliography{nablachem}

\begin{thebibliography}{71}%
\makeatletter
\providecommand \@ifxundefined [1]{%
 \@ifx{#1\undefined}
}%
\providecommand \@ifnum [1]{%
 \ifnum #1\expandafter \@firstoftwo
 \else \expandafter \@secondoftwo
 \fi
}%
\providecommand \@ifx [1]{%
 \ifx #1\expandafter \@firstoftwo
 \else \expandafter \@secondoftwo
 \fi
}%
\providecommand \natexlab [1]{#1}%
\providecommand \enquote  [1]{``#1''}%
\providecommand \bibnamefont  [1]{#1}%
\providecommand \bibfnamefont [1]{#1}%
\providecommand \citenamefont [1]{#1}%
\providecommand \href@noop [0]{\@secondoftwo}%
\providecommand \href [0]{\begingroup \@sanitize@url \@href}%
\providecommand \@href[1]{\@@startlink{#1}\@@href}%
\providecommand \@@href[1]{\endgroup#1\@@endlink}%
\providecommand \@sanitize@url [0]{\catcode `\\12\catcode `\$12\catcode `\&12\catcode `\#12\catcode `\^12\catcode `\_12\catcode `\%12\relax}%
\providecommand \@@startlink[1]{}%
\providecommand \@@endlink[0]{}%
\providecommand \url  [0]{\begingroup\@sanitize@url \@url }%
\providecommand \@url [1]{\endgroup\@href {#1}{\urlprefix }}%
\providecommand \urlprefix  [0]{URL }%
\providecommand \Eprint [0]{\href }%
\providecommand \doibase [0]{https://doi.org/}%
\providecommand \selectlanguage [0]{\@gobble}%
\providecommand \bibinfo  [0]{\@secondoftwo}%
\providecommand \bibfield  [0]{\@secondoftwo}%
\providecommand \translation [1]{[#1]}%
\providecommand \BibitemOpen [0]{}%
\providecommand \bibitemStop [0]{}%
\providecommand \bibitemNoStop [0]{.\EOS\space}%
\providecommand \EOS [0]{\spacefactor3000\relax}%
\providecommand \BibitemShut  [1]{\csname bibitem#1\endcsname}%
\let\auto@bib@innerbib\@empty
\bibitem [{\citenamefont {Keith}\ \emph {et~al.}(2021)\citenamefont {Keith}, \citenamefont {Vassilev-Galindo}, \citenamefont {Cheng}, \citenamefont {Chmiela}, \citenamefont {Gastegger}, \citenamefont {Müller},\ and\ \citenamefont {Tkatchenko}}]{keith2021combining}%
  \BibitemOpen
  \bibfield  {author} {\bibinfo {author} {\bibfnamefont {J.~A.}\ \bibnamefont {Keith}}, \bibinfo {author} {\bibfnamefont {V.}~\bibnamefont {Vassilev-Galindo}}, \bibinfo {author} {\bibfnamefont {B.}~\bibnamefont {Cheng}}, \bibinfo {author} {\bibfnamefont {S.}~\bibnamefont {Chmiela}}, \bibinfo {author} {\bibfnamefont {M.}~\bibnamefont {Gastegger}}, \bibinfo {author} {\bibfnamefont {K.-R.}\ \bibnamefont {Müller}},\ and\ \bibinfo {author} {\bibfnamefont {A.}~\bibnamefont {Tkatchenko}},\ }\bibfield  {title} {\bibinfo {title} {Combining machine learning and computational chemistry for predictive modeling and design},\ }\href {https://doi.org/10.1021/acs.chemrev.1c00107} {\bibfield  {journal} {\bibinfo  {journal} {Chemical Reviews}\ }\textbf {\bibinfo {volume} {121}},\ \bibinfo {pages} {9816} (\bibinfo {year} {2021})},\ \bibinfo {note} {publisher: American Chemical Society}\BibitemShut {NoStop}%
\bibitem [{\citenamefont {Lu}\ \emph {et~al.}(2024)\citenamefont {Lu}, \citenamefont {Xia}, \citenamefont {Ren}, \citenamefont {Xie}, \citenamefont {Zhou}, \citenamefont {Vinai}, \citenamefont {Morton}, \citenamefont {Wee}, \citenamefont {van~der Wiel}, \citenamefont {Zhang},\ and\ \citenamefont {Wong}}]{Lu2024}%
  \BibitemOpen
  \bibfield  {author} {\bibinfo {author} {\bibfnamefont {B.}~\bibnamefont {Lu}}, \bibinfo {author} {\bibfnamefont {Y.}~\bibnamefont {Xia}}, \bibinfo {author} {\bibfnamefont {Y.}~\bibnamefont {Ren}}, \bibinfo {author} {\bibfnamefont {M.}~\bibnamefont {Xie}}, \bibinfo {author} {\bibfnamefont {L.}~\bibnamefont {Zhou}}, \bibinfo {author} {\bibfnamefont {G.}~\bibnamefont {Vinai}}, \bibinfo {author} {\bibfnamefont {S.~A.}\ \bibnamefont {Morton}}, \bibinfo {author} {\bibfnamefont {A.~T.~S.}\ \bibnamefont {Wee}}, \bibinfo {author} {\bibfnamefont {W.~G.}\ \bibnamefont {van~der Wiel}}, \bibinfo {author} {\bibfnamefont {W.}~\bibnamefont {Zhang}},\ and\ \bibinfo {author} {\bibfnamefont {P.~K.~J.}\ \bibnamefont {Wong}},\ }\bibfield  {title} {\bibinfo {title} {When machine learning meets {2D} materials: a review},\ }\bibfield  {journal} {\bibinfo  {journal} {Advanced Science}\ }\textbf {\bibinfo {volume} {11}},\ \href {https://doi.org/10.1002/advs.202305277} {10.1002/advs.202305277} (\bibinfo {year} {2024}),\ \bibinfo
  {note} {publisher: Wiley}\BibitemShut {NoStop}%
\bibitem [{\citenamefont {Von~Lilienfeld}\ \emph {et~al.}(2020)\citenamefont {Von~Lilienfeld}, \citenamefont {Müller},\ and\ \citenamefont {Tkatchenko}}]{VonLilienfeld2020}%
  \BibitemOpen
  \bibfield  {author} {\bibinfo {author} {\bibfnamefont {O.~A.}\ \bibnamefont {Von~Lilienfeld}}, \bibinfo {author} {\bibfnamefont {K.-R.}\ \bibnamefont {Müller}},\ and\ \bibinfo {author} {\bibfnamefont {A.}~\bibnamefont {Tkatchenko}},\ }\bibfield  {title} {\bibinfo {title} {Exploring chemical compound space with quantum-based machine learning},\ }\href {https://doi.org/10.1038/s41570-020-0189-9} {\bibfield  {journal} {\bibinfo  {journal} {Nature Reviews Chemistry}\ }\textbf {\bibinfo {volume} {4}},\ \bibinfo {pages} {347} (\bibinfo {year} {2020})}\BibitemShut {NoStop}%
\bibitem [{\citenamefont {Smith}\ \emph {et~al.}(2017)\citenamefont {Smith}, \citenamefont {Isayev},\ and\ \citenamefont {Roitberg}}]{Smith2017a}%
  \BibitemOpen
  \bibfield  {author} {\bibinfo {author} {\bibfnamefont {J.~S.}\ \bibnamefont {Smith}}, \bibinfo {author} {\bibfnamefont {O.}~\bibnamefont {Isayev}},\ and\ \bibinfo {author} {\bibfnamefont {A.~E.}\ \bibnamefont {Roitberg}},\ }\bibfield  {title} {\bibinfo {title} {{ANI}-1, {A} data set of 20 million calculated off-equilibrium conformations for organic molecules},\ }\href {https://doi.org/10.1038/sdata.2017.193} {\bibfield  {journal} {\bibinfo  {journal} {Scientific Data}\ }\textbf {\bibinfo {volume} {4}},\ \bibinfo {pages} {170193} (\bibinfo {year} {2017})}\BibitemShut {NoStop}%
\bibitem [{\citenamefont {Ganscha}\ \emph {et~al.}(2025)\citenamefont {Ganscha}, \citenamefont {Unke}, \citenamefont {Ahlin}, \citenamefont {Maennel}, \citenamefont {Kashubin},\ and\ \citenamefont {Müller}}]{Ganscha2025}%
  \BibitemOpen
  \bibfield  {author} {\bibinfo {author} {\bibfnamefont {S.}~\bibnamefont {Ganscha}}, \bibinfo {author} {\bibfnamefont {O.~T.}\ \bibnamefont {Unke}}, \bibinfo {author} {\bibfnamefont {D.}~\bibnamefont {Ahlin}}, \bibinfo {author} {\bibfnamefont {H.}~\bibnamefont {Maennel}}, \bibinfo {author} {\bibfnamefont {S.}~\bibnamefont {Kashubin}},\ and\ \bibinfo {author} {\bibfnamefont {K.-R.}\ \bibnamefont {Müller}},\ }\bibfield  {title} {\bibinfo {title} {The {QCML} dataset, {Quantum} chemistry reference data from 33.{5M} {DFT} and 14.{7B} semi-empirical calculations},\ }\href {https://doi.org/10.1038/s41597-025-04720-7} {\bibfield  {journal} {\bibinfo  {journal} {Scientific Data}\ }\textbf {\bibinfo {volume} {12}},\ \bibinfo {pages} {406} (\bibinfo {year} {2025})}\BibitemShut {NoStop}%
\bibitem [{\citenamefont {Madych}\ and\ \citenamefont {Nelson}(1992)}]{Madych1992}%
  \BibitemOpen
  \bibfield  {author} {\bibinfo {author} {\bibfnamefont {W.}~\bibnamefont {Madych}}\ and\ \bibinfo {author} {\bibfnamefont {S.}~\bibnamefont {Nelson}},\ }\bibfield  {title} {\bibinfo {title} {Bounds on multivariate polynomials and exponential error estimates for multiquadric interpolation},\ }\href {https://doi.org/10.1016/0021-9045(92)90058-v} {\bibfield  {journal} {\bibinfo  {journal} {Journal of Approximation Theory}\ }\textbf {\bibinfo {volume} {70}},\ \bibinfo {pages} {94} (\bibinfo {year} {1992})},\ \bibinfo {note} {publisher: Elsevier BV}\BibitemShut {NoStop}%
\bibitem [{\citenamefont {Krenn}\ \emph {et~al.}(2022)\citenamefont {Krenn}, \citenamefont {Ai}, \citenamefont {Barthel}, \citenamefont {Carson}, \citenamefont {Frei}, \citenamefont {Frey}, \citenamefont {Friederich}, \citenamefont {Gaudin}, \citenamefont {Gayle}, \citenamefont {Jablonka}, \citenamefont {Lameiro}, \citenamefont {Lemm}, \citenamefont {Lo}, \citenamefont {Moosavi}, \citenamefont {Nápoles-Duarte}, \citenamefont {Nigam}, \citenamefont {Pollice}, \citenamefont {Rajan}, \citenamefont {Schatzschneider}, \citenamefont {Schwaller}, \citenamefont {Skreta}, \citenamefont {Smit}, \citenamefont {Strieth-Kalthoff}, \citenamefont {Sun}, \citenamefont {Tom}, \citenamefont {von Rudorff}, \citenamefont {Wang}, \citenamefont {White}, \citenamefont {Young}, \citenamefont {Yu},\ and\ \citenamefont {Aspuru-Guzik}}]{Krenn2022}%
  \BibitemOpen
  \bibfield  {author} {\bibinfo {author} {\bibfnamefont {M.}~\bibnamefont {Krenn}}, \bibinfo {author} {\bibfnamefont {Q.}~\bibnamefont {Ai}}, \bibinfo {author} {\bibfnamefont {S.}~\bibnamefont {Barthel}}, \bibinfo {author} {\bibfnamefont {N.}~\bibnamefont {Carson}}, \bibinfo {author} {\bibfnamefont {A.}~\bibnamefont {Frei}}, \bibinfo {author} {\bibfnamefont {N.~C.}\ \bibnamefont {Frey}}, \bibinfo {author} {\bibfnamefont {P.}~\bibnamefont {Friederich}}, \bibinfo {author} {\bibfnamefont {T.}~\bibnamefont {Gaudin}}, \bibinfo {author} {\bibfnamefont {A.~A.}\ \bibnamefont {Gayle}}, \bibinfo {author} {\bibfnamefont {K.~M.}\ \bibnamefont {Jablonka}}, \bibinfo {author} {\bibfnamefont {R.~F.}\ \bibnamefont {Lameiro}}, \bibinfo {author} {\bibfnamefont {D.}~\bibnamefont {Lemm}}, \bibinfo {author} {\bibfnamefont {A.}~\bibnamefont {Lo}}, \bibinfo {author} {\bibfnamefont {S.~M.}\ \bibnamefont {Moosavi}}, \bibinfo {author} {\bibfnamefont {J.~M.}\ \bibnamefont {Nápoles-Duarte}}, \bibinfo {author} {\bibfnamefont
  {A.}~\bibnamefont {Nigam}}, \bibinfo {author} {\bibfnamefont {R.}~\bibnamefont {Pollice}}, \bibinfo {author} {\bibfnamefont {K.}~\bibnamefont {Rajan}}, \bibinfo {author} {\bibfnamefont {U.}~\bibnamefont {Schatzschneider}}, \bibinfo {author} {\bibfnamefont {P.}~\bibnamefont {Schwaller}}, \bibinfo {author} {\bibfnamefont {M.}~\bibnamefont {Skreta}}, \bibinfo {author} {\bibfnamefont {B.}~\bibnamefont {Smit}}, \bibinfo {author} {\bibfnamefont {F.}~\bibnamefont {Strieth-Kalthoff}}, \bibinfo {author} {\bibfnamefont {C.}~\bibnamefont {Sun}}, \bibinfo {author} {\bibfnamefont {G.}~\bibnamefont {Tom}}, \bibinfo {author} {\bibfnamefont {G.~F.}\ \bibnamefont {von Rudorff}}, \bibinfo {author} {\bibfnamefont {A.}~\bibnamefont {Wang}}, \bibinfo {author} {\bibfnamefont {A.~D.}\ \bibnamefont {White}}, \bibinfo {author} {\bibfnamefont {A.}~\bibnamefont {Young}}, \bibinfo {author} {\bibfnamefont {R.}~\bibnamefont {Yu}},\ and\ \bibinfo {author} {\bibfnamefont {A.}~\bibnamefont {Aspuru-Guzik}},\ }\bibfield  {title} {\bibinfo
  {title} {{SELFIES} and the future of molecular string representations},\ }\href {https://doi.org/10.1016/j.patter.2022.100588} {\bibfield  {journal} {\bibinfo  {journal} {Patterns}\ }\textbf {\bibinfo {volume} {3}},\ \bibinfo {pages} {100588} (\bibinfo {year} {2022})},\ \bibinfo {note} {publisher: Elsevier BV}\BibitemShut {NoStop}%
\bibitem [{\citenamefont {Musil}\ \emph {et~al.}(2021)\citenamefont {Musil}, \citenamefont {Grisafi}, \citenamefont {Bartók}, \citenamefont {Ortner}, \citenamefont {Csányi},\ and\ \citenamefont {Ceriotti}}]{Musil2021}%
  \BibitemOpen
  \bibfield  {author} {\bibinfo {author} {\bibfnamefont {F.}~\bibnamefont {Musil}}, \bibinfo {author} {\bibfnamefont {A.}~\bibnamefont {Grisafi}}, \bibinfo {author} {\bibfnamefont {A.~P.}\ \bibnamefont {Bartók}}, \bibinfo {author} {\bibfnamefont {C.}~\bibnamefont {Ortner}}, \bibinfo {author} {\bibfnamefont {G.}~\bibnamefont {Csányi}},\ and\ \bibinfo {author} {\bibfnamefont {M.}~\bibnamefont {Ceriotti}},\ }\bibfield  {title} {\bibinfo {title} {Physics-{Inspired} {Structural} {Representations} for {Molecules} and {Materials}},\ }\href {https://doi.org/10.1021/acs.chemrev.1c00021} {\bibfield  {journal} {\bibinfo  {journal} {Chemical Reviews}\ }\textbf {\bibinfo {volume} {121}},\ \bibinfo {pages} {9759} (\bibinfo {year} {2021})}\BibitemShut {NoStop}%
\bibitem [{\citenamefont {David}\ \emph {et~al.}(2020)\citenamefont {David}, \citenamefont {Thakkar}, \citenamefont {Mercado},\ and\ \citenamefont {Engkvist}}]{David2020}%
  \BibitemOpen
  \bibfield  {author} {\bibinfo {author} {\bibfnamefont {L.}~\bibnamefont {David}}, \bibinfo {author} {\bibfnamefont {A.}~\bibnamefont {Thakkar}}, \bibinfo {author} {\bibfnamefont {R.}~\bibnamefont {Mercado}},\ and\ \bibinfo {author} {\bibfnamefont {O.}~\bibnamefont {Engkvist}},\ }\bibfield  {title} {\bibinfo {title} {Molecular representations in {AI}-driven drug discovery: a review and practical guide},\ }\href {https://doi.org/10.1186/s13321-020-00460-5} {\bibfield  {journal} {\bibinfo  {journal} {Journal of Cheminformatics}\ }\textbf {\bibinfo {volume} {12}},\ \bibinfo {pages} {56} (\bibinfo {year} {2020})}\BibitemShut {NoStop}%
\bibitem [{\citenamefont {Briling}\ \emph {et~al.}(2024)\citenamefont {Briling}, \citenamefont {Calvino~Alonso}, \citenamefont {Fabrizio},\ and\ \citenamefont {Corminboeuf}}]{Briling2024}%
  \BibitemOpen
  \bibfield  {author} {\bibinfo {author} {\bibfnamefont {K.~R.}\ \bibnamefont {Briling}}, \bibinfo {author} {\bibfnamefont {Y.}~\bibnamefont {Calvino~Alonso}}, \bibinfo {author} {\bibfnamefont {A.}~\bibnamefont {Fabrizio}},\ and\ \bibinfo {author} {\bibfnamefont {C.}~\bibnamefont {Corminboeuf}},\ }\bibfield  {title} {\bibinfo {title} {{SPA}$^{\textrm{{h}}}$ {M}(a,b): {Encoding} the {Density} {Information} from {Guess} {Hamiltonian} in {Quantum} {Machine} {Learning} {Representations}},\ }\href {https://doi.org/10.1021/acs.jctc.3c01040} {\bibfield  {journal} {\bibinfo  {journal} {Journal of Chemical Theory and Computation}\ }\textbf {\bibinfo {volume} {20}},\ \bibinfo {pages} {1108} (\bibinfo {year} {2024})}\BibitemShut {NoStop}%
\bibitem [{\citenamefont {Karandashev}\ and\ \citenamefont {Von~Lilienfeld}(2022)}]{Karandashev2022}%
  \BibitemOpen
  \bibfield  {author} {\bibinfo {author} {\bibfnamefont {K.}~\bibnamefont {Karandashev}}\ and\ \bibinfo {author} {\bibfnamefont {O.~A.}\ \bibnamefont {Von~Lilienfeld}},\ }\bibfield  {title} {\bibinfo {title} {An orbital-based representation for accurate quantum machine learning},\ }\href {https://doi.org/10.1063/5.0083301} {\bibfield  {journal} {\bibinfo  {journal} {The Journal of Chemical Physics}\ }\textbf {\bibinfo {volume} {156}},\ \bibinfo {pages} {114101} (\bibinfo {year} {2022})}\BibitemShut {NoStop}%
\bibitem [{\citenamefont {Faber}\ \emph {et~al.}(2018)\citenamefont {Faber}, \citenamefont {Christensen}, \citenamefont {Huang},\ and\ \citenamefont {von Lilienfeld}}]{Faber2018}%
  \BibitemOpen
  \bibfield  {author} {\bibinfo {author} {\bibfnamefont {F.~A.}\ \bibnamefont {Faber}}, \bibinfo {author} {\bibfnamefont {A.~S.}\ \bibnamefont {Christensen}}, \bibinfo {author} {\bibfnamefont {B.}~\bibnamefont {Huang}},\ and\ \bibinfo {author} {\bibfnamefont {O.~A.}\ \bibnamefont {von Lilienfeld}},\ }\bibfield  {title} {\bibinfo {title} {Alchemical and structural distribution based representation for universal quantum machine learning},\ }\href {https://doi.org/10.1063/1.5020710} {\bibfield  {journal} {\bibinfo  {journal} {The Journal of Chemical Physics}\ }\textbf {\bibinfo {volume} {148}},\ \bibinfo {pages} {241717} (\bibinfo {year} {2018})},\ \bibinfo {note} {publisher: AIP Publishing tex.timestamp: 2019-11-24}\BibitemShut {NoStop}%
\bibitem [{\citenamefont {Khan}\ \emph {et~al.}(2023)\citenamefont {Khan}, \citenamefont {Heinen},\ and\ \citenamefont {Von~Lilienfeld}}]{Khan2023}%
  \BibitemOpen
  \bibfield  {author} {\bibinfo {author} {\bibfnamefont {D.}~\bibnamefont {Khan}}, \bibinfo {author} {\bibfnamefont {S.}~\bibnamefont {Heinen}},\ and\ \bibinfo {author} {\bibfnamefont {O.~A.}\ \bibnamefont {Von~Lilienfeld}},\ }\bibfield  {title} {\bibinfo {title} {Kernel based quantum machine learning at record rate: {Many}-body distribution functionals as compact representations},\ }\href {https://doi.org/10.1063/5.0152215} {\bibfield  {journal} {\bibinfo  {journal} {The Journal of Chemical Physics}\ }\textbf {\bibinfo {volume} {159}},\ \bibinfo {pages} {034106} (\bibinfo {year} {2023})}\BibitemShut {NoStop}%
\bibitem [{\citenamefont {Weinreich}\ \emph {et~al.}(2023)\citenamefont {Weinreich}, \citenamefont {von Rudorff},\ and\ \citenamefont {von Lilienfeld}}]{Weinreich2023a}%
  \BibitemOpen
  \bibfield  {author} {\bibinfo {author} {\bibfnamefont {J.}~\bibnamefont {Weinreich}}, \bibinfo {author} {\bibfnamefont {G.~F.}\ \bibnamefont {von Rudorff}},\ and\ \bibinfo {author} {\bibfnamefont {O.~A.}\ \bibnamefont {von Lilienfeld}},\ }\bibfield  {title} {\bibinfo {title} {Encrypted machine learning of molecular quantum properties},\ }\href {https://doi.org/10.1088/2632-2153/acc928} {\bibfield  {journal} {\bibinfo  {journal} {Machine Learning: Science and Technology}\ }\textbf {\bibinfo {volume} {4}},\ \bibinfo {pages} {025017} (\bibinfo {year} {2023})},\ \bibinfo {note} {publisher: IOP Publishing}\BibitemShut {NoStop}%
\bibitem [{\citenamefont {Jia}\ \emph {et~al.}(2021)\citenamefont {Jia}, \citenamefont {Sun}, \citenamefont {Lian},\ and\ \citenamefont {Hou}}]{jia2021feature}%
  \BibitemOpen
  \bibfield  {author} {\bibinfo {author} {\bibfnamefont {W.}~\bibnamefont {Jia}}, \bibinfo {author} {\bibfnamefont {M.}~\bibnamefont {Sun}}, \bibinfo {author} {\bibfnamefont {J.}~\bibnamefont {Lian}},\ and\ \bibinfo {author} {\bibfnamefont {S.}~\bibnamefont {Hou}},\ }\bibfield  {title} {\bibinfo {title} {Feature dimensionality reduction: a review},\ }\href {https://doi.org/10.1007/s40747-021-00637-x} {\bibfield  {journal} {\bibinfo  {journal} {Complex \& Intelligent Systems}\ }\textbf {\bibinfo {volume} {8}},\ \bibinfo {pages} {2663} (\bibinfo {year} {2021})},\ \bibinfo {note} {publisher: Springer}\BibitemShut {NoStop}%
\bibitem [{\citenamefont {van~der Maaten}\ \emph {et~al.}(2009)\citenamefont {van~der Maaten}, \citenamefont {Postma},\ and\ \citenamefont {van~den Herik}}]{vandermaaten2009dimensionality}%
  \BibitemOpen
  \bibfield  {author} {\bibinfo {author} {\bibfnamefont {L.}~\bibnamefont {van~der Maaten}}, \bibinfo {author} {\bibfnamefont {E.}~\bibnamefont {Postma}},\ and\ \bibinfo {author} {\bibfnamefont {J.}~\bibnamefont {van~den Herik}},\ }\bibfield  {title} {\bibinfo {title} {Dimensionality reduction: a comparative review},\ }\href@noop {} {\bibfield  {journal} {\bibinfo  {journal} {Journal of Machine Learning Research}\ }\textbf {\bibinfo {volume} {10}},\ \bibinfo {pages} {66} (\bibinfo {year} {2009})},\ \bibinfo {note} {publisher: MIT Press}\BibitemShut {NoStop}%
\bibitem [{\citenamefont {Camastra}\ and\ \citenamefont {Staiano}(2016)}]{camastra2016intrinsic}%
  \BibitemOpen
  \bibfield  {author} {\bibinfo {author} {\bibfnamefont {F.}~\bibnamefont {Camastra}}\ and\ \bibinfo {author} {\bibfnamefont {A.}~\bibnamefont {Staiano}},\ }\bibfield  {title} {\bibinfo {title} {Intrinsic dimension estimation: {Advances} and open problems},\ }\href {https://doi.org/10.1016/j.ins.2015.08.029} {\bibfield  {journal} {\bibinfo  {journal} {Information Sciences}\ }\textbf {\bibinfo {volume} {328}},\ \bibinfo {pages} {26} (\bibinfo {year} {2016})},\ \bibinfo {note} {publisher: Elsevier}\BibitemShut {NoStop}%
\bibitem [{\citenamefont {Orlov}\ \emph {et~al.}(2025)\citenamefont {Orlov}, \citenamefont {Akhmetshin}, \citenamefont {Horvath}, \citenamefont {Marcou},\ and\ \citenamefont {Varnek}}]{orlov2025exploring}%
  \BibitemOpen
  \bibfield  {author} {\bibinfo {author} {\bibfnamefont {A.~A.}\ \bibnamefont {Orlov}}, \bibinfo {author} {\bibfnamefont {T.~N.}\ \bibnamefont {Akhmetshin}}, \bibinfo {author} {\bibfnamefont {D.}~\bibnamefont {Horvath}}, \bibinfo {author} {\bibfnamefont {G.}~\bibnamefont {Marcou}},\ and\ \bibinfo {author} {\bibfnamefont {A.}~\bibnamefont {Varnek}},\ }\bibfield  {title} {\bibinfo {title} {From high dimensions to human insight: {Exploring} dimensionality reduction for chemical space visualization},\ }\href {https://doi.org/10.1002/minf.202400265} {\bibfield  {journal} {\bibinfo  {journal} {Molecular Informatics}\ }\textbf {\bibinfo {volume} {44}},\ \bibinfo {pages} {e202400265} (\bibinfo {year} {2025})},\ \bibinfo {note} {publisher: Wiley-VCH}\BibitemShut {NoStop}%
\bibitem [{\citenamefont {Tamilselvi}\ and\ \citenamefont {Chandrasekaran}(2017)}]{tamilselvi2017review}%
  \BibitemOpen
  \bibfield  {author} {\bibinfo {author} {\bibfnamefont {J.}~\bibnamefont {Tamilselvi}}\ and\ \bibinfo {author} {\bibfnamefont {R.~M.}\ \bibnamefont {Chandrasekaran}},\ }\bibfield  {title} {\bibinfo {title} {A review on dimensionality reduction for machine learning},\ }\href@noop {} {\bibfield  {journal} {\bibinfo  {journal} {International Journal of Computer Applications}\ }\textbf {\bibinfo {volume} {173}},\ \bibinfo {pages} {42} (\bibinfo {year} {2017})},\ \bibinfo {note} {publisher: Foundation of Computer Science}\BibitemShut {NoStop}%
\bibitem [{\citenamefont {Williams}\ and\ \citenamefont {Ilieş}(2018)}]{Williams2018}%
  \BibitemOpen
  \bibfield  {author} {\bibinfo {author} {\bibfnamefont {R.~M.}\ \bibnamefont {Williams}}\ and\ \bibinfo {author} {\bibfnamefont {H.~T.}\ \bibnamefont {Ilieş}},\ }\bibfield  {title} {\bibinfo {title} {Practical shape analysis and segmentation methods for point cloud models},\ }\href {https://doi.org/10.1016/j.cviu.2018.07.006} {\bibfield  {journal} {\bibinfo  {journal} {Computer Vision and Image Understanding}\ }\textbf {\bibinfo {volume} {174}},\ \bibinfo {pages} {28} (\bibinfo {year} {2018})},\ \bibinfo {note} {publisher: Elsevier}\BibitemShut {NoStop}%
\bibitem [{\citenamefont {Laikov}(2011)}]{Laikov2011}%
  \BibitemOpen
  \bibfield  {author} {\bibinfo {author} {\bibfnamefont {D.~N.}\ \bibnamefont {Laikov}},\ }\bibfield  {title} {\bibinfo {title} {Intrinsic minimal atomic basis representation of molecular electronic wavefunctions},\ }\href {https://doi.org/10.1002/qua.22767} {\bibfield  {journal} {\bibinfo  {journal} {International Journal of Quantum Chemistry}\ }\textbf {\bibinfo {volume} {111}},\ \bibinfo {pages} {2851} (\bibinfo {year} {2011})}\BibitemShut {NoStop}%
\bibitem [{\citenamefont {Schleder}\ \emph {et~al.}(2019)\citenamefont {Schleder}, \citenamefont {Padilha}, \citenamefont {Acosta}, \citenamefont {Costa},\ and\ \citenamefont {Fazzio}}]{Schleder2019}%
  \BibitemOpen
  \bibfield  {author} {\bibinfo {author} {\bibfnamefont {G.~R.}\ \bibnamefont {Schleder}}, \bibinfo {author} {\bibfnamefont {A.~C.~M.}\ \bibnamefont {Padilha}}, \bibinfo {author} {\bibfnamefont {C.~M.}\ \bibnamefont {Acosta}}, \bibinfo {author} {\bibfnamefont {M.}~\bibnamefont {Costa}},\ and\ \bibinfo {author} {\bibfnamefont {A.}~\bibnamefont {Fazzio}},\ }\bibfield  {title} {\bibinfo {title} {From {DFT} to machine learning: recent approaches to materials science a review},\ }\href {https://doi.org/10.1088/2515-7639/ab084b} {\bibfield  {journal} {\bibinfo  {journal} {Journal of Physics: Materials}\ }\textbf {\bibinfo {volume} {2}},\ \bibinfo {pages} {032001} (\bibinfo {year} {2019})},\ \bibinfo {note} {publisher: IOP Publishing}\BibitemShut {NoStop}%
\bibitem [{\citenamefont {Ceruti}\ \emph {et~al.}(2014)\citenamefont {Ceruti}, \citenamefont {Bassis}, \citenamefont {Rozza}, \citenamefont {Lombardi}, \citenamefont {Casiraghi},\ and\ \citenamefont {Campadelli}}]{Ceruti2014}%
  \BibitemOpen
  \bibfield  {author} {\bibinfo {author} {\bibfnamefont {C.}~\bibnamefont {Ceruti}}, \bibinfo {author} {\bibfnamefont {S.}~\bibnamefont {Bassis}}, \bibinfo {author} {\bibfnamefont {A.}~\bibnamefont {Rozza}}, \bibinfo {author} {\bibfnamefont {G.}~\bibnamefont {Lombardi}}, \bibinfo {author} {\bibfnamefont {E.}~\bibnamefont {Casiraghi}},\ and\ \bibinfo {author} {\bibfnamefont {P.}~\bibnamefont {Campadelli}},\ }\bibfield  {title} {\bibinfo {title} {{DANCo}: {An} intrinsic dimensionality estimator exploiting angle and norm concentration},\ }\href {https://doi.org/10.1016/j.patcog.2014.02.013} {\bibfield  {journal} {\bibinfo  {journal} {Pattern Recognition}\ }\textbf {\bibinfo {volume} {47}},\ \bibinfo {pages} {2569} (\bibinfo {year} {2014})}\BibitemShut {NoStop}%
\bibitem [{\citenamefont {Albergante}\ \emph {et~al.}(2019)\citenamefont {Albergante}, \citenamefont {Bac},\ and\ \citenamefont {Zinovyev}}]{Albergante2019}%
  \BibitemOpen
  \bibfield  {author} {\bibinfo {author} {\bibfnamefont {L.}~\bibnamefont {Albergante}}, \bibinfo {author} {\bibfnamefont {J.}~\bibnamefont {Bac}},\ and\ \bibinfo {author} {\bibfnamefont {A.}~\bibnamefont {Zinovyev}},\ }\bibfield  {title} {\bibinfo {title} {Estimating the effective dimension of large biological datasets using {Fisher} separability analysis},\ }in\ \href {https://doi.org/10.1109/IJCNN.2019.8852450} {\emph {\bibinfo {booktitle} {2019 {International} {Joint} {Conference} on {Neural} {Networks} ({IJCNN})}}}\ (\bibinfo  {publisher} {IEEE},\ \bibinfo {address} {Budapest, Hungary},\ \bibinfo {year} {2019})\ pp.\ \bibinfo {pages} {1--8}\BibitemShut {NoStop}%
\bibitem [{\citenamefont {Facco}\ \emph {et~al.}(2017)\citenamefont {Facco}, \citenamefont {d’Errico}, \citenamefont {Rodriguez},\ and\ \citenamefont {Laio}}]{Facco2017}%
  \BibitemOpen
  \bibfield  {author} {\bibinfo {author} {\bibfnamefont {E.}~\bibnamefont {Facco}}, \bibinfo {author} {\bibfnamefont {M.}~\bibnamefont {d’Errico}}, \bibinfo {author} {\bibfnamefont {A.}~\bibnamefont {Rodriguez}},\ and\ \bibinfo {author} {\bibfnamefont {A.}~\bibnamefont {Laio}},\ }\bibfield  {title} {\bibinfo {title} {Estimating the intrinsic dimension of datasets by a minimal neighborhood information},\ }\href {https://doi.org/10.1038/s41598-017-11873-y} {\bibfield  {journal} {\bibinfo  {journal} {Scientific Reports}\ }\textbf {\bibinfo {volume} {7}},\ \bibinfo {pages} {12140} (\bibinfo {year} {2017})}\BibitemShut {NoStop}%
\bibitem [{\citenamefont {Denti}\ \emph {et~al.}(2022)\citenamefont {Denti}, \citenamefont {Doimo}, \citenamefont {Laio},\ and\ \citenamefont {Mira}}]{Denti2022}%
  \BibitemOpen
  \bibfield  {author} {\bibinfo {author} {\bibfnamefont {F.}~\bibnamefont {Denti}}, \bibinfo {author} {\bibfnamefont {D.}~\bibnamefont {Doimo}}, \bibinfo {author} {\bibfnamefont {A.}~\bibnamefont {Laio}},\ and\ \bibinfo {author} {\bibfnamefont {A.}~\bibnamefont {Mira}},\ }\bibfield  {title} {\bibinfo {title} {The generalized ratios intrinsic dimension estimator},\ }\href {https://doi.org/10.1038/s41598-022-20991-1} {\bibfield  {journal} {\bibinfo  {journal} {Scientific Reports}\ }\textbf {\bibinfo {volume} {12}},\ \bibinfo {pages} {20005} (\bibinfo {year} {2022})}\BibitemShut {NoStop}%
\bibitem [{\citenamefont {Gomtsyan}\ \emph {et~al.}(2019)\citenamefont {Gomtsyan}, \citenamefont {Mokrov}, \citenamefont {Panov},\ and\ \citenamefont {Yanovich}}]{Gomtsyan2019}%
  \BibitemOpen
  \bibfield  {author} {\bibinfo {author} {\bibfnamefont {M.}~\bibnamefont {Gomtsyan}}, \bibinfo {author} {\bibfnamefont {N.}~\bibnamefont {Mokrov}}, \bibinfo {author} {\bibfnamefont {M.}~\bibnamefont {Panov}},\ and\ \bibinfo {author} {\bibfnamefont {Y.}~\bibnamefont {Yanovich}},\ }\bibfield  {title} {\bibinfo {title} {Geometry-{Aware} {Maximum} {Likelihood} {Estimation} of {Intrinsic} {Dimension}},\ }in\ \href {https://proceedings.mlr.press/v101/gomtsyan19a.html} {\emph {\bibinfo {booktitle} {Proceedings of {The} {Eleventh} {Asian} {Conference} on {Machine} {Learning}}}}\ (\bibinfo  {publisher} {PMLR},\ \bibinfo {year} {2019})\ pp.\ \bibinfo {pages} {1126--1141},\ \bibinfo {note} {iSSN: 2640-3498}\BibitemShut {NoStop}%
\bibitem [{\citenamefont {Bouveyron}\ \emph {et~al.}(2011)\citenamefont {Bouveyron}, \citenamefont {Celeux},\ and\ \citenamefont {Girard}}]{Bouveyron2011}%
  \BibitemOpen
  \bibfield  {author} {\bibinfo {author} {\bibfnamefont {C.}~\bibnamefont {Bouveyron}}, \bibinfo {author} {\bibfnamefont {G.}~\bibnamefont {Celeux}},\ and\ \bibinfo {author} {\bibfnamefont {S.}~\bibnamefont {Girard}},\ }\bibfield  {title} {\bibinfo {title} {Intrinsic dimension estimation by maximum likelihood in isotropic probabilistic {PCA}},\ }\href {https://doi.org/10.1016/j.patrec.2011.07.017} {\bibfield  {journal} {\bibinfo  {journal} {Pattern Recognition Letters}\ }\textbf {\bibinfo {volume} {32}},\ \bibinfo {pages} {1706} (\bibinfo {year} {2011})}\BibitemShut {NoStop}%
\bibitem [{\citenamefont {Rozza}\ \emph {et~al.}(2012)\citenamefont {Rozza}, \citenamefont {Lombardi}, \citenamefont {Ceruti}, \citenamefont {Casiraghi},\ and\ \citenamefont {Campadelli}}]{Rozza2012}%
  \BibitemOpen
  \bibfield  {author} {\bibinfo {author} {\bibfnamefont {A.}~\bibnamefont {Rozza}}, \bibinfo {author} {\bibfnamefont {G.}~\bibnamefont {Lombardi}}, \bibinfo {author} {\bibfnamefont {C.}~\bibnamefont {Ceruti}}, \bibinfo {author} {\bibfnamefont {E.}~\bibnamefont {Casiraghi}},\ and\ \bibinfo {author} {\bibfnamefont {P.}~\bibnamefont {Campadelli}},\ }\bibfield  {title} {\bibinfo {title} {Novel high intrinsic dimensionality estimators},\ }\href {https://doi.org/10.1007/s10994-012-5294-7} {\bibfield  {journal} {\bibinfo  {journal} {Machine Learning}\ }\textbf {\bibinfo {volume} {89}},\ \bibinfo {pages} {37} (\bibinfo {year} {2012})}\BibitemShut {NoStop}%
\bibitem [{\citenamefont {Erba}\ \emph {et~al.}(2019)\citenamefont {Erba}, \citenamefont {Gherardi},\ and\ \citenamefont {Rotondo}}]{Erba2019}%
  \BibitemOpen
  \bibfield  {author} {\bibinfo {author} {\bibfnamefont {V.}~\bibnamefont {Erba}}, \bibinfo {author} {\bibfnamefont {M.}~\bibnamefont {Gherardi}},\ and\ \bibinfo {author} {\bibfnamefont {P.}~\bibnamefont {Rotondo}},\ }\bibfield  {title} {\bibinfo {title} {Intrinsic dimension estimation for locally undersampled data},\ }\href {https://doi.org/10.1038/s41598-019-53549-9} {\bibfield  {journal} {\bibinfo  {journal} {Scientific Reports}\ }\textbf {\bibinfo {volume} {9}},\ \bibinfo {pages} {17133} (\bibinfo {year} {2019})}\BibitemShut {NoStop}%
\bibitem [{\citenamefont {Grunwald}\ \emph {et~al.}(2015)\citenamefont {Grunwald}, \citenamefont {Basak},\ and\ \citenamefont {Basak}}]{Grunwald2015}%
  \BibitemOpen
  \bibfield  {author} {\bibinfo {author} {\bibfnamefont {G.}~\bibnamefont {Grunwald}}, \bibinfo {author} {\bibfnamefont {S.}~\bibnamefont {Basak}},\ and\ \bibinfo {author} {\bibfnamefont {S.}~\bibnamefont {Basak}},\ }\bibfield  {title} {\bibinfo {title} {Intrinsic dimensionality of chemical space: {Characterization} and applications},\ }in\ \href {https://doi.org/10.3390/MOL2NET-1-b037} {\emph {\bibinfo {booktitle} {Proceedings of {MOL2NET}, {International} {Conference} on {Multidisciplinary} {Sciences}}}}\ (\bibinfo  {publisher} {MDPI},\ \bibinfo {address} {Sciforum.net},\ \bibinfo {year} {2015})\ p.\ \bibinfo {pages} {b037}\BibitemShut {NoStop}%
\bibitem [{\citenamefont {Shukur}\ \emph {et~al.}(2011)\citenamefont {Shukur}, \citenamefont {Rani}, \citenamefont {Bhavani}, \citenamefont {Sastry},\ and\ \citenamefont {Raju}}]{Shukur2011}%
  \BibitemOpen
  \bibfield  {author} {\bibinfo {author} {\bibfnamefont {M.~H.}\ \bibnamefont {Shukur}}, \bibinfo {author} {\bibfnamefont {T.~S.}\ \bibnamefont {Rani}}, \bibinfo {author} {\bibfnamefont {S.~D.}\ \bibnamefont {Bhavani}}, \bibinfo {author} {\bibfnamefont {G.~N.}\ \bibnamefont {Sastry}},\ and\ \bibinfo {author} {\bibfnamefont {S.~B.}\ \bibnamefont {Raju}},\ }\bibfield  {title} {\bibinfo {title} {Local and {Global} {Intrinsic} {Dimensionality} {Estimation} for {Better} {Chemical} {Space} {Representation}},\ }in\ \href {https://doi.org/10.1007/978-3-642-25725-4_29} {\emph {\bibinfo {booktitle} {Multi-disciplinary {Trends} in {Artificial} {Intelligence}}}},\ Vol.\ \bibinfo {volume} {7080},\ \bibinfo {editor} {edited by\ \bibinfo {editor} {\bibfnamefont {C.}~\bibnamefont {Sombattheera}}, \bibinfo {editor} {\bibfnamefont {A.}~\bibnamefont {Agarwal}}, \bibinfo {editor} {\bibfnamefont {S.~K.}\ \bibnamefont {Udgata}},\ and\ \bibinfo {editor} {\bibfnamefont {K.}~\bibnamefont {Lavangnananda}}}\ (\bibinfo  {publisher}
  {Springer Berlin Heidelberg},\ \bibinfo {address} {Berlin, Heidelberg},\ \bibinfo {year} {2011})\ pp.\ \bibinfo {pages} {329--338},\ \bibinfo {note} {series Title: Lecture Notes in Computer Science}\BibitemShut {NoStop}%
\bibitem [{\citenamefont {Le}\ and\ \citenamefont {Winkler}(2018)}]{Le2018}%
  \BibitemOpen
  \bibfield  {author} {\bibinfo {author} {\bibfnamefont {T.~C.}\ \bibnamefont {Le}}\ and\ \bibinfo {author} {\bibfnamefont {D.~A.}\ \bibnamefont {Winkler}},\ }\bibfield  {title} {\bibinfo {title} {Applications in {Materials} {Science}},\ }in\ \href {https://doi.org/10.1002/9783527806539.ch12} {\emph {\bibinfo {booktitle} {Applied {Chemoinformatics}}}},\ \bibinfo {editor} {edited by\ \bibinfo {editor} {\bibfnamefont {T.}~\bibnamefont {Engel}}\ and\ \bibinfo {editor} {\bibfnamefont {J.}~\bibnamefont {Gasteiger}}}\ (\bibinfo  {publisher} {Wiley},\ \bibinfo {year} {2018})\ \bibinfo {edition} {1st}\ ed.,\ pp.\ \bibinfo {pages} {547--569}\BibitemShut {NoStop}%
\bibitem [{\citenamefont {Yao}\ \emph {et~al.}(2022)\citenamefont {Yao}, \citenamefont {Zhang}, \citenamefont {Hu}, \citenamefont {Chang}, \citenamefont {Liu},\ and\ \citenamefont {Zhao}}]{Yao2022}%
  \BibitemOpen
  \bibfield  {author} {\bibinfo {author} {\bibfnamefont {X.}~\bibnamefont {Yao}}, \bibinfo {author} {\bibfnamefont {R.}~\bibnamefont {Zhang}}, \bibinfo {author} {\bibfnamefont {J.}~\bibnamefont {Hu}}, \bibinfo {author} {\bibfnamefont {K.}~\bibnamefont {Chang}}, \bibinfo {author} {\bibfnamefont {X.}~\bibnamefont {Liu}},\ and\ \bibinfo {author} {\bibfnamefont {J.}~\bibnamefont {Zhao}},\ }\bibfield  {title} {\bibinfo {title} {Combining intrinsic dimension and local tangent space for manifold spectral clustering image segmentation},\ }\href {https://doi.org/10.1007/s00500-022-06751-3} {\bibfield  {journal} {\bibinfo  {journal} {Soft Computing}\ }\textbf {\bibinfo {volume} {26}},\ \bibinfo {pages} {9557} (\bibinfo {year} {2022})}\BibitemShut {NoStop}%
\bibitem [{\citenamefont {von Rudorff}\ and\ \citenamefont {von Lilienfeld}(2020)}]{von_Rudorff_2020}%
  \BibitemOpen
  \bibfield  {author} {\bibinfo {author} {\bibfnamefont {G.~F.}\ \bibnamefont {von Rudorff}}\ and\ \bibinfo {author} {\bibfnamefont {O.~A.}\ \bibnamefont {von Lilienfeld}},\ }\bibfield  {title} {\bibinfo {title} {Alchemical perturbation density functional theory},\ }\href {https://doi.org/10.1103/physrevresearch.2.023220} {\bibfield  {journal} {\bibinfo  {journal} {Physical Review Research}\ }\textbf {\bibinfo {volume} {2}},\ \bibinfo {pages} {023220} (\bibinfo {year} {2020})},\ \bibinfo {note} {publisher: American Physical Society (APS)}\BibitemShut {NoStop}%
\bibitem [{\citenamefont {von Rudorff}(2021)}]{Rudorff2021a}%
  \BibitemOpen
  \bibfield  {author} {\bibinfo {author} {\bibfnamefont {G.~F.}\ \bibnamefont {von Rudorff}},\ }\bibfield  {title} {\bibinfo {title} {Arbitrarily accurate quantum alchemy},\ }\href {https://doi.org/10.1063/5.0073941} {\bibfield  {journal} {\bibinfo  {journal} {The Journal of Chemical Physics}\ ,\ \bibinfo {pages} {224103}} (\bibinfo {year} {2021})},\ \bibinfo {note} {publisher: AIP Publishing}\BibitemShut {NoStop}%
\bibitem [{\citenamefont {von Lilienfeld}(2009)}]{Lilienfeld2009}%
  \BibitemOpen
  \bibfield  {author} {\bibinfo {author} {\bibfnamefont {O.~A.}\ \bibnamefont {von Lilienfeld}},\ }\bibfield  {title} {\bibinfo {title} {Accurate ab initio energy gradients in chemical compound space},\ }\href {https://doi.org/10.1063/1.3249969} {\bibfield  {journal} {\bibinfo  {journal} {Journal of Chemical Physics}\ }\textbf {\bibinfo {volume} {131}},\ \bibinfo {pages} {164102} (\bibinfo {year} {2009})},\ \bibinfo {note} {publisher: AIP Publishing tex.timestamp: 2018-08-23}\BibitemShut {NoStop}%
\bibitem [{\citenamefont {Fias}\ \emph {et~al.}(2018)\citenamefont {Fias}, \citenamefont {Chang},\ and\ \citenamefont {von Lilienfeld}}]{Fias2018}%
  \BibitemOpen
  \bibfield  {author} {\bibinfo {author} {\bibfnamefont {S.}~\bibnamefont {Fias}}, \bibinfo {author} {\bibfnamefont {S.}~\bibnamefont {Chang}},\ and\ \bibinfo {author} {\bibfnamefont {O.~A.}\ \bibnamefont {von Lilienfeld}},\ }\bibfield  {title} {\bibinfo {title} {Alchemical normal modes unify chemical space},\ }\bibfield  {journal} {\bibinfo  {journal} {Journal of Physical Chemistry Letters}\ }\href {https://doi.org/10.1021/acs.jpclett.8b02805} {10.1021/acs.jpclett.8b02805} (\bibinfo {year} {2018}),\ \bibinfo {note} {publisher: American Chemical Society (ACS) tex.timestamp: 2018-12-13}\BibitemShut {NoStop}%
\bibitem [{\citenamefont {Wilson}(1962)}]{Wilson1962}%
  \BibitemOpen
  \bibfield  {author} {\bibinfo {author} {\bibfnamefont {E.~B.}\ \bibnamefont {Wilson}},\ }\bibfield  {title} {\bibinfo {title} {Four-dimensional electron density function},\ }\href {https://doi.org/10.1063/1.1732864} {\bibfield  {journal} {\bibinfo  {journal} {Journal of Chemical Physics}\ }\textbf {\bibinfo {volume} {36}},\ \bibinfo {pages} {2232} (\bibinfo {year} {1962})},\ \bibinfo {note} {publisher: AIP Publishing tex.timestamp: 2018-08-20}\BibitemShut {NoStop}%
\bibitem [{\citenamefont {Ayers}\ \emph {et~al.}(2009)\citenamefont {Ayers}, \citenamefont {Liu},\ and\ \citenamefont {Li}}]{Ayers2009}%
  \BibitemOpen
  \bibfield  {author} {\bibinfo {author} {\bibfnamefont {P.~W.}\ \bibnamefont {Ayers}}, \bibinfo {author} {\bibfnamefont {S.}~\bibnamefont {Liu}},\ and\ \bibinfo {author} {\bibfnamefont {T.}~\bibnamefont {Li}},\ }\bibfield  {title} {\bibinfo {title} {Chargephilicity and chargephobicity: {Two} new reactivity indicators for external potential changes from density functional reactivity theory},\ }\href {https://doi.org/10.1016/j.cplett.2009.08.067} {\bibfield  {journal} {\bibinfo  {journal} {Chemical Physics Letters}\ }\textbf {\bibinfo {volume} {480}},\ \bibinfo {pages} {318} (\bibinfo {year} {2009})},\ \bibinfo {note} {publisher: Elsevier BV}\BibitemShut {NoStop}%
\bibitem [{\citenamefont {Balawender}\ \emph {et~al.}(2013)\citenamefont {Balawender}, \citenamefont {Welearegay}, \citenamefont {Lesiuk}, \citenamefont {De~Proft},\ and\ \citenamefont {Geerlings}}]{Balawender2013}%
  \BibitemOpen
  \bibfield  {author} {\bibinfo {author} {\bibfnamefont {R.}~\bibnamefont {Balawender}}, \bibinfo {author} {\bibfnamefont {M.~A.}\ \bibnamefont {Welearegay}}, \bibinfo {author} {\bibfnamefont {M.}~\bibnamefont {Lesiuk}}, \bibinfo {author} {\bibfnamefont {F.}~\bibnamefont {De~Proft}},\ and\ \bibinfo {author} {\bibfnamefont {P.}~\bibnamefont {Geerlings}},\ }\bibfield  {title} {\bibinfo {title} {Exploring chemical space with the alchemical derivatives},\ }\href {https://doi.org/10.1021/ct400706g} {\bibfield  {journal} {\bibinfo  {journal} {Journal of Chemical Theory and Computation}\ }\textbf {\bibinfo {volume} {9}},\ \bibinfo {pages} {5327} (\bibinfo {year} {2013})},\ \bibinfo {note} {publisher: American Chemical Society (ACS)}\BibitemShut {NoStop}%
\bibitem [{\citenamefont {Lesiuk}\ \emph {et~al.}(2012)\citenamefont {Lesiuk}, \citenamefont {Balawender},\ and\ \citenamefont {Zachara}}]{Lesiuk2012}%
  \BibitemOpen
  \bibfield  {author} {\bibinfo {author} {\bibfnamefont {M.}~\bibnamefont {Lesiuk}}, \bibinfo {author} {\bibfnamefont {R.}~\bibnamefont {Balawender}},\ and\ \bibinfo {author} {\bibfnamefont {J.}~\bibnamefont {Zachara}},\ }\bibfield  {title} {\bibinfo {title} {Higher order alchemical derivatives from coupled perturbed self-consistent field theory},\ }\href {https://doi.org/10.1063/1.3674163} {\bibfield  {journal} {\bibinfo  {journal} {The Journal of Chemical Physics}\ }\textbf {\bibinfo {volume} {136}},\ \bibinfo {pages} {034104} (\bibinfo {year} {2012})},\ \bibinfo {note} {publisher: AIP Publishing}\BibitemShut {NoStop}%
\bibitem [{\citenamefont {Tamayo-Mendoza}\ \emph {et~al.}(2018)\citenamefont {Tamayo-Mendoza}, \citenamefont {Kreisbeck}, \citenamefont {Lindh},\ and\ \citenamefont {Aspuru-Guzik}}]{TamayoMendoza2018}%
  \BibitemOpen
  \bibfield  {author} {\bibinfo {author} {\bibfnamefont {T.}~\bibnamefont {Tamayo-Mendoza}}, \bibinfo {author} {\bibfnamefont {C.}~\bibnamefont {Kreisbeck}}, \bibinfo {author} {\bibfnamefont {R.}~\bibnamefont {Lindh}},\ and\ \bibinfo {author} {\bibfnamefont {A.}~\bibnamefont {Aspuru-Guzik}},\ }\bibfield  {title} {\bibinfo {title} {Automatic differentiation in quantum chemistry with applications to fully variational hartree–fock},\ }\href {https://doi.org/10.1021/acscentsci.7b00586} {\bibfield  {journal} {\bibinfo  {journal} {ACS Central Science}\ }\textbf {\bibinfo {volume} {4}},\ \bibinfo {pages} {559} (\bibinfo {year} {2018})},\ \bibinfo {note} {publisher: American Chemical Society (ACS)}\BibitemShut {NoStop}%
\bibitem [{\citenamefont {von Lilienfeld}\ and\ \citenamefont {Tuckerman}(2006)}]{Lilienfeld2006}%
  \BibitemOpen
  \bibfield  {author} {\bibinfo {author} {\bibfnamefont {O.~A.}\ \bibnamefont {von Lilienfeld}}\ and\ \bibinfo {author} {\bibfnamefont {M.~E.}\ \bibnamefont {Tuckerman}},\ }\bibfield  {title} {\bibinfo {title} {Molecular grand-canonical ensemble density functional theory and exploration of chemical space},\ }\href {https://doi.org/10.1063/1.2338537} {\bibfield  {journal} {\bibinfo  {journal} {The Journal of Chemical Physics}\ }\textbf {\bibinfo {volume} {125}},\ \bibinfo {pages} {154104} (\bibinfo {year} {2006})},\ \bibinfo {note} {publisher: AIP Publishing}\BibitemShut {NoStop}%
\bibitem [{\citenamefont {Griego}\ \emph {et~al.}(2018)\citenamefont {Griego}, \citenamefont {Saravanan},\ and\ \citenamefont {Keith}}]{Griego2018}%
  \BibitemOpen
  \bibfield  {author} {\bibinfo {author} {\bibfnamefont {C.~D.}\ \bibnamefont {Griego}}, \bibinfo {author} {\bibfnamefont {K.}~\bibnamefont {Saravanan}},\ and\ \bibinfo {author} {\bibfnamefont {J.~A.}\ \bibnamefont {Keith}},\ }\bibfield  {title} {\bibinfo {title} {Benchmarking computational alchemy for carbide, nitride, and oxide catalysts},\ }\href {https://doi.org/10.1002/adts.201800142} {\bibfield  {journal} {\bibinfo  {journal} {Advanced Theory and Simulations}\ }\textbf {\bibinfo {volume} {2}},\ \bibinfo {pages} {1800142} (\bibinfo {year} {2018})},\ \bibinfo {note} {publisher: Wiley}\BibitemShut {NoStop}%
\bibitem [{\citenamefont {Kasim}\ \emph {et~al.}(2022)\citenamefont {Kasim}, \citenamefont {Lehtola},\ and\ \citenamefont {Vinko}}]{Kasim_2022}%
  \BibitemOpen
  \bibfield  {author} {\bibinfo {author} {\bibfnamefont {M.~F.}\ \bibnamefont {Kasim}}, \bibinfo {author} {\bibfnamefont {S.}~\bibnamefont {Lehtola}},\ and\ \bibinfo {author} {\bibfnamefont {S.~M.}\ \bibnamefont {Vinko}},\ }\bibfield  {title} {\bibinfo {title} {{DQC}: {A} {Python} program package for differentiable quantum chemistry},\ }\href {https://doi.org/10.1063/5.0076202} {\bibfield  {journal} {\bibinfo  {journal} {The Journal of Chemical Physics}\ }\textbf {\bibinfo {volume} {156}},\ \bibinfo {pages} {084801} (\bibinfo {year} {2022})},\ \bibinfo {note} {publisher: AIP Publishing}\BibitemShut {NoStop}%
\bibitem [{\citenamefont {Muñoz}\ \emph {et~al.}(2020)\citenamefont {Muñoz}, \citenamefont {Robles-Navarro}, \citenamefont {Fuentealba},\ and\ \citenamefont {Cárdenas}}]{Munoz2020}%
  \BibitemOpen
  \bibfield  {author} {\bibinfo {author} {\bibfnamefont {M.}~\bibnamefont {Muñoz}}, \bibinfo {author} {\bibfnamefont {A.}~\bibnamefont {Robles-Navarro}}, \bibinfo {author} {\bibfnamefont {P.}~\bibnamefont {Fuentealba}},\ and\ \bibinfo {author} {\bibfnamefont {C.}~\bibnamefont {Cárdenas}},\ }\bibfield  {title} {\bibinfo {title} {Predicting deprotonation sites using alchemical derivatives},\ }\href {https://doi.org/10.1021/acs.jpca.9b09472} {\bibfield  {journal} {\bibinfo  {journal} {The Journal of Physical Chemistry A}\ }\textbf {\bibinfo {volume} {124}},\ \bibinfo {pages} {3754} (\bibinfo {year} {2020})},\ \bibinfo {note} {publisher: American Chemical Society (ACS) tex.timestamp: 2020-08-21}\BibitemShut {NoStop}%
\bibitem [{\citenamefont {Miranda-Quintana}\ and\ \citenamefont {Ayers}(2017)}]{MirandaQuintana2017}%
  \BibitemOpen
  \bibfield  {author} {\bibinfo {author} {\bibfnamefont {R.~A.}\ \bibnamefont {Miranda-Quintana}}\ and\ \bibinfo {author} {\bibfnamefont {P.~W.}\ \bibnamefont {Ayers}},\ }\bibfield  {title} {\bibinfo {title} {Interpolating {Hamiltonians} in chemical compound space},\ }\href {https://doi.org/10.1002/qua.25384} {\bibfield  {journal} {\bibinfo  {journal} {International Journal of Quantum Chemistry}\ }\textbf {\bibinfo {volume} {117}},\ \bibinfo {pages} {e25384} (\bibinfo {year} {2017})},\ \bibinfo {note} {publisher: Wiley}\BibitemShut {NoStop}%
\bibitem [{\citenamefont {Abbott}\ \emph {et~al.}(2021)\citenamefont {Abbott}, \citenamefont {Abbott}, \citenamefont {Turney},\ and\ \citenamefont {Schaefer}}]{Abbott2021}%
  \BibitemOpen
  \bibfield  {author} {\bibinfo {author} {\bibfnamefont {A.~S.}\ \bibnamefont {Abbott}}, \bibinfo {author} {\bibfnamefont {B.~Z.}\ \bibnamefont {Abbott}}, \bibinfo {author} {\bibfnamefont {J.~M.}\ \bibnamefont {Turney}},\ and\ \bibinfo {author} {\bibfnamefont {H.~F.}\ \bibnamefont {Schaefer}},\ }\bibfield  {title} {\bibinfo {title} {Arbitrary-order derivatives of quantum chemical methods via automatic differentiation},\ }\href {https://doi.org/10.1021/acs.jpclett.1c00607} {\bibfield  {journal} {\bibinfo  {journal} {The Journal of Physical Chemistry Letters}\ }\textbf {\bibinfo {volume} {12}},\ \bibinfo {pages} {3232} (\bibinfo {year} {2021})},\ \bibinfo {note} {publisher: American Chemical Society (ACS)}\BibitemShut {NoStop}%
\bibitem [{\citenamefont {Perdew}\ \emph {et~al.}(1996)\citenamefont {Perdew}, \citenamefont {Burke},\ and\ \citenamefont {Ernzerhof}}]{Perdew1996}%
  \BibitemOpen
  \bibfield  {author} {\bibinfo {author} {\bibfnamefont {J.~P.}\ \bibnamefont {Perdew}}, \bibinfo {author} {\bibfnamefont {K.}~\bibnamefont {Burke}},\ and\ \bibinfo {author} {\bibfnamefont {M.}~\bibnamefont {Ernzerhof}},\ }\bibfield  {title} {\bibinfo {title} {Generalized {Gradient} {Approximation} {Made} {Simple}},\ }\href {https://doi.org/10.1103/PhysRevLett.77.3865} {\bibfield  {journal} {\bibinfo  {journal} {Physical Review Letters}\ }\textbf {\bibinfo {volume} {77}},\ \bibinfo {pages} {3865} (\bibinfo {year} {1996})}\BibitemShut {NoStop}%
\bibitem [{\citenamefont {Sun}\ \emph {et~al.}(2017)\citenamefont {Sun}, \citenamefont {Berkelbach}, \citenamefont {Blunt}, \citenamefont {Booth}, \citenamefont {Guo}, \citenamefont {Li}, \citenamefont {Liu}, \citenamefont {McClain}, \citenamefont {Sayfutyarova}, \citenamefont {Sharma}, \citenamefont {Wouters},\ and\ \citenamefont {Chan}}]{PYSCF}%
  \BibitemOpen
  \bibfield  {author} {\bibinfo {author} {\bibfnamefont {Q.}~\bibnamefont {Sun}}, \bibinfo {author} {\bibfnamefont {T.~C.}\ \bibnamefont {Berkelbach}}, \bibinfo {author} {\bibfnamefont {N.~S.}\ \bibnamefont {Blunt}}, \bibinfo {author} {\bibfnamefont {G.~H.}\ \bibnamefont {Booth}}, \bibinfo {author} {\bibfnamefont {S.}~\bibnamefont {Guo}}, \bibinfo {author} {\bibfnamefont {Z.}~\bibnamefont {Li}}, \bibinfo {author} {\bibfnamefont {J.}~\bibnamefont {Liu}}, \bibinfo {author} {\bibfnamefont {J.~D.}\ \bibnamefont {McClain}}, \bibinfo {author} {\bibfnamefont {E.~R.}\ \bibnamefont {Sayfutyarova}}, \bibinfo {author} {\bibfnamefont {S.}~\bibnamefont {Sharma}}, \bibinfo {author} {\bibfnamefont {S.}~\bibnamefont {Wouters}},\ and\ \bibinfo {author} {\bibfnamefont {G.~K.}\ \bibnamefont {Chan}},\ }\href {https://doi.org/10.1002/wcms.1340} {\bibinfo {title} {{PySCF}: the {Python}‐based simulations of chemistry framework}} (\bibinfo {year} {2017}),\ \bibinfo {note} {number: 1 Pages: e1340 Volume: 8 tex.eprint:
  https://onlinelibrary.wiley.com/doi/pdf/10.1002/wcms.1340 tex.timestamp: 2019-11-24}\BibitemShut {NoStop}%
\bibitem [{\citenamefont {Domenichini}\ \emph {et~al.}(2020)\citenamefont {Domenichini}, \citenamefont {von Rudorff},\ and\ \citenamefont {von Lilienfeld}}]{Domenichini2020}%
  \BibitemOpen
  \bibfield  {author} {\bibinfo {author} {\bibfnamefont {G.}~\bibnamefont {Domenichini}}, \bibinfo {author} {\bibfnamefont {G.~F.}\ \bibnamefont {von Rudorff}},\ and\ \bibinfo {author} {\bibfnamefont {O.~A.}\ \bibnamefont {von Lilienfeld}},\ }\bibfield  {title} {\bibinfo {title} {Effects of perturbation order and basis set on alchemical predictions},\ }\href {https://doi.org/10.1063/5.0023590} {\bibfield  {journal} {\bibinfo  {journal} {The Journal of Chemical Physics}\ }\textbf {\bibinfo {volume} {153}},\ \bibinfo {pages} {144118} (\bibinfo {year} {2020})},\ \bibinfo {note} {publisher: AIP Publishing}\BibitemShut {NoStop}%
\bibitem [{\citenamefont {Domenichini}\ and\ \citenamefont {von Lilienfeld}(2022)}]{Domenichini2022}%
  \BibitemOpen
  \bibfield  {author} {\bibinfo {author} {\bibfnamefont {G.}~\bibnamefont {Domenichini}}\ and\ \bibinfo {author} {\bibfnamefont {O.~A.}\ \bibnamefont {von Lilienfeld}},\ }\bibfield  {title} {\bibinfo {title} {Alchemical geometry relaxation},\ }\href {https://doi.org/10.1063/5.0085817} {\bibfield  {journal} {\bibinfo  {journal} {The Journal of Chemical Physics}\ }\textbf {\bibinfo {volume} {156}},\ \bibinfo {pages} {184801} (\bibinfo {year} {2022})},\ \bibinfo {note} {publisher: AIP Publishing}\BibitemShut {NoStop}%
\bibitem [{\citenamefont {Zhang}\ and\ \citenamefont {Chan}(2022)}]{Zhang2022}%
  \BibitemOpen
  \bibfield  {author} {\bibinfo {author} {\bibfnamefont {X.}~\bibnamefont {Zhang}}\ and\ \bibinfo {author} {\bibfnamefont {G.~K.-L.}\ \bibnamefont {Chan}},\ }\bibfield  {title} {\bibinfo {title} {Differentiable quantum chemistry with {PySCF} for molecules and materials at the mean-field level and beyond},\ }\href {https://doi.org/10.1063/5.0118200} {\bibfield  {journal} {\bibinfo  {journal} {The Journal of Chemical Physics}\ }\textbf {\bibinfo {volume} {157}},\ \bibinfo {pages} {204801} (\bibinfo {year} {2022})},\ \bibinfo {note} {publisher: AIP Publishing}\BibitemShut {NoStop}%
\bibitem [{\citenamefont {Osamura}\ \emph {et~al.}(1986)\citenamefont {Osamura}, \citenamefont {Yamaguchi},\ and\ \citenamefont {Schaefer}}]{Osamura1986}%
  \BibitemOpen
  \bibfield  {author} {\bibinfo {author} {\bibfnamefont {Y.}~\bibnamefont {Osamura}}, \bibinfo {author} {\bibfnamefont {Y.}~\bibnamefont {Yamaguchi}},\ and\ \bibinfo {author} {\bibfnamefont {H.~F.}\ \bibnamefont {Schaefer}},\ }\bibfield  {title} {\bibinfo {title} {Second-order coupled perturbed hartree—fock equations for closed-shell and open-shell self-consistent-field wavefunctions},\ }\href {https://doi.org/10.1016/0301-0104(86)80023-4} {\bibfield  {journal} {\bibinfo  {journal} {Chemical Physics}\ }\textbf {\bibinfo {volume} {103}},\ \bibinfo {pages} {227} (\bibinfo {year} {1986})}\BibitemShut {NoStop}%
\bibitem [{\citenamefont {Zhou}\ \emph {et~al.}(2023)\citenamefont {Zhou}, \citenamefont {Zhou}, \citenamefont {Hua}, \citenamefont {Bawane},\ and\ \citenamefont {Feng}}]{Zhou2023}%
  \BibitemOpen
  \bibfield  {author} {\bibinfo {author} {\bibfnamefont {H.}~\bibnamefont {Zhou}}, \bibinfo {author} {\bibfnamefont {S.}~\bibnamefont {Zhou}}, \bibinfo {author} {\bibfnamefont {Z.}~\bibnamefont {Hua}}, \bibinfo {author} {\bibfnamefont {K.}~\bibnamefont {Bawane}},\ and\ \bibinfo {author} {\bibfnamefont {T.}~\bibnamefont {Feng}},\ }\bibfield  {title} {\bibinfo {title} {Extreme sensitivity of higher-order interatomic force constants and thermal conductivity to the energy surface roughness of exchange-correlation functionals},\ }\href {https://doi.org/10.1063/5.0173762} {\bibfield  {journal} {\bibinfo  {journal} {Applied Physics Letters}\ }\textbf {\bibinfo {volume} {123}},\ \bibinfo {pages} {192201} (\bibinfo {year} {2023})}\BibitemShut {NoStop}%
\bibitem [{\citenamefont {Banjafar}\ and\ \citenamefont {von Rudorff}(2025{\natexlab{a}})}]{Banjafar2025}%
  \BibitemOpen
  \bibfield  {author} {\bibinfo {author} {\bibfnamefont {A.}~\bibnamefont {Banjafar}}\ and\ \bibinfo {author} {\bibfnamefont {G.~F.}\ \bibnamefont {von Rudorff}},\ }\href {https://doi.org/10.5281/ZENODO.15789764} {\bibinfo {title} {Intrinsic {Dimensionality} of {Molecular} {Properties}}} (\bibinfo {year} {2025}{\natexlab{a}})\BibitemShut {NoStop}%
\bibitem [{\citenamefont {Mendez}\ \emph {et~al.}(2019)\citenamefont {Mendez}, \citenamefont {Gaulton}, \citenamefont {Bento}, \citenamefont {Chambers}, \citenamefont {De Veij}, \citenamefont {Félix}, \citenamefont {Magariños}, \citenamefont {Mosquera}, \citenamefont {Mutowo}, \citenamefont {Nowotka}, \citenamefont {Gordillo-Marañón}, \citenamefont {Hunter}, \citenamefont {Junco}, \citenamefont {Mugumbate}, \citenamefont {Rodriguez-Lopez}, \citenamefont {Atkinson}, \citenamefont {Bosc}, \citenamefont {Radoux}, \citenamefont {Segura-Cabrera}, \citenamefont {Hersey},\ and\ \citenamefont {Leach}}]{Mendez2019}%
  \BibitemOpen
  \bibfield  {author} {\bibinfo {author} {\bibfnamefont {D.}~\bibnamefont {Mendez}}, \bibinfo {author} {\bibfnamefont {A.}~\bibnamefont {Gaulton}}, \bibinfo {author} {\bibfnamefont {A.~P.}\ \bibnamefont {Bento}}, \bibinfo {author} {\bibfnamefont {J.}~\bibnamefont {Chambers}}, \bibinfo {author} {\bibfnamefont {M.}~\bibnamefont {De Veij}}, \bibinfo {author} {\bibfnamefont {E.}~\bibnamefont {Félix}}, \bibinfo {author} {\bibfnamefont {M.}~\bibnamefont {Magariños}}, \bibinfo {author} {\bibfnamefont {J.}~\bibnamefont {Mosquera}}, \bibinfo {author} {\bibfnamefont {P.}~\bibnamefont {Mutowo}}, \bibinfo {author} {\bibfnamefont {M.}~\bibnamefont {Nowotka}}, \bibinfo {author} {\bibfnamefont {M.}~\bibnamefont {Gordillo-Marañón}}, \bibinfo {author} {\bibfnamefont {F.}~\bibnamefont {Hunter}}, \bibinfo {author} {\bibfnamefont {L.}~\bibnamefont {Junco}}, \bibinfo {author} {\bibfnamefont {G.}~\bibnamefont {Mugumbate}}, \bibinfo {author} {\bibfnamefont {M.}~\bibnamefont {Rodriguez-Lopez}}, \bibinfo {author} {\bibfnamefont
  {F.}~\bibnamefont {Atkinson}}, \bibinfo {author} {\bibfnamefont {N.}~\bibnamefont {Bosc}}, \bibinfo {author} {\bibfnamefont {C.}~\bibnamefont {Radoux}}, \bibinfo {author} {\bibfnamefont {A.}~\bibnamefont {Segura-Cabrera}}, \bibinfo {author} {\bibfnamefont {A.}~\bibnamefont {Hersey}},\ and\ \bibinfo {author} {\bibfnamefont {A.}~\bibnamefont {Leach}},\ }\bibfield  {title} {\bibinfo {title} {{ChEMBL}: towards direct deposition of bioassay data},\ }\href {https://doi.org/10.1093/nar/gky1075} {\bibfield  {journal} {\bibinfo  {journal} {Nucleic Acids Research}\ }\textbf {\bibinfo {volume} {47}},\ \bibinfo {pages} {D930} (\bibinfo {year} {2019})}\BibitemShut {NoStop}%
\bibitem [{\citenamefont {Fyodorov}\ and\ \citenamefont {Le~Doussal}(2018)}]{Fyodorov2018}%
  \BibitemOpen
  \bibfield  {author} {\bibinfo {author} {\bibfnamefont {Y.~V.}\ \bibnamefont {Fyodorov}}\ and\ \bibinfo {author} {\bibfnamefont {P.}~\bibnamefont {Le~Doussal}},\ }\bibfield  {title} {\bibinfo {title} {Hessian spectrum at the global minimum of high-dimensional random landscapes},\ }\href {https://doi.org/10.1088/1751-8121/aae74f} {\bibfield  {journal} {\bibinfo  {journal} {Journal of Physics A: Mathematical and Theoretical}\ }\textbf {\bibinfo {volume} {51}},\ \bibinfo {pages} {474002} (\bibinfo {year} {2018})}\BibitemShut {NoStop}%
\bibitem [{\citenamefont {Dreßler}\ and\ \citenamefont {Sebastiani}(2020)}]{Dressler2020}%
  \BibitemOpen
  \bibfield  {author} {\bibinfo {author} {\bibfnamefont {C.}~\bibnamefont {Dreßler}}\ and\ \bibinfo {author} {\bibfnamefont {D.}~\bibnamefont {Sebastiani}},\ }\bibfield  {title} {\bibinfo {title} {Reduced eigensystem representation of the linear density‐density response function},\ }\href {https://doi.org/10.1002/qua.26085} {\bibfield  {journal} {\bibinfo  {journal} {International Journal of Quantum Chemistry}\ }\textbf {\bibinfo {volume} {120}},\ \bibinfo {pages} {e26085} (\bibinfo {year} {2020})}\BibitemShut {NoStop}%
\bibitem [{\citenamefont {De~Hoop}\ and\ \citenamefont {Prange}(2007)}]{DeHoop2007}%
  \BibitemOpen
  \bibfield  {author} {\bibinfo {author} {\bibfnamefont {A.~T.}\ \bibnamefont {De~Hoop}}\ and\ \bibinfo {author} {\bibfnamefont {M.~D.}\ \bibnamefont {Prange}},\ }\bibfield  {title} {\bibinfo {title} {Variational analysis of the natural decay rates and eigenmodes of cavity-enclosed diffusive fields},\ }\href {https://doi.org/10.1088/1751-8113/40/41/014} {\bibfield  {journal} {\bibinfo  {journal} {Journal of Physics A: Mathematical and Theoretical}\ }\textbf {\bibinfo {volume} {40}},\ \bibinfo {pages} {12463} (\bibinfo {year} {2007})}\BibitemShut {NoStop}%
\bibitem [{\citenamefont {Shawe-Taylor}\ \emph {et~al.}(2005)\citenamefont {Shawe-Taylor}, \citenamefont {Williams}, \citenamefont {Cristianini},\ and\ \citenamefont {Kandola}}]{Shawe-Taylor2005}%
  \BibitemOpen
  \bibfield  {author} {\bibinfo {author} {\bibfnamefont {J.}~\bibnamefont {Shawe-Taylor}}, \bibinfo {author} {\bibfnamefont {C.}~\bibnamefont {Williams}}, \bibinfo {author} {\bibfnamefont {N.}~\bibnamefont {Cristianini}},\ and\ \bibinfo {author} {\bibfnamefont {J.}~\bibnamefont {Kandola}},\ }\bibfield  {title} {\bibinfo {title} {On the {Eigenspectrum} of the {Gram} {Matrix} and the {Generalization} {Error} of {Kernel}-{PCA}},\ }\href {https://doi.org/10.1109/TIT.2005.850052} {\bibfield  {journal} {\bibinfo  {journal} {IEEE Transactions on Information Theory}\ }\textbf {\bibinfo {volume} {51}},\ \bibinfo {pages} {2510} (\bibinfo {year} {2005})}\BibitemShut {NoStop}%
\bibitem [{\citenamefont {Stöhr}\ \emph {et~al.}(2019)\citenamefont {Stöhr}, \citenamefont {Van~Voorhis},\ and\ \citenamefont {Tkatchenko}}]{Stohr2019}%
  \BibitemOpen
  \bibfield  {author} {\bibinfo {author} {\bibfnamefont {M.}~\bibnamefont {Stöhr}}, \bibinfo {author} {\bibfnamefont {T.}~\bibnamefont {Van~Voorhis}},\ and\ \bibinfo {author} {\bibfnamefont {A.}~\bibnamefont {Tkatchenko}},\ }\bibfield  {title} {\bibinfo {title} {Theory and practice of modeling van der {Waals} interactions in electronic-structure calculations},\ }\href {https://doi.org/10.1039/C9CS00060G} {\bibfield  {journal} {\bibinfo  {journal} {Chemical Society Reviews}\ }\textbf {\bibinfo {volume} {48}},\ \bibinfo {pages} {4118} (\bibinfo {year} {2019})}\BibitemShut {NoStop}%
\bibitem [{\citenamefont {Fedik}\ \emph {et~al.}(2022)\citenamefont {Fedik}, \citenamefont {Zubatyuk}, \citenamefont {Kulichenko}, \citenamefont {Lubbers}, \citenamefont {Smith}, \citenamefont {Nebgen}, \citenamefont {Messerly}, \citenamefont {Li}, \citenamefont {Boldyrev}, \citenamefont {Barros}, \citenamefont {Isayev},\ and\ \citenamefont {Tretiak}}]{Fedik2022}%
  \BibitemOpen
  \bibfield  {author} {\bibinfo {author} {\bibfnamefont {N.}~\bibnamefont {Fedik}}, \bibinfo {author} {\bibfnamefont {R.}~\bibnamefont {Zubatyuk}}, \bibinfo {author} {\bibfnamefont {M.}~\bibnamefont {Kulichenko}}, \bibinfo {author} {\bibfnamefont {N.}~\bibnamefont {Lubbers}}, \bibinfo {author} {\bibfnamefont {J.~S.}\ \bibnamefont {Smith}}, \bibinfo {author} {\bibfnamefont {B.}~\bibnamefont {Nebgen}}, \bibinfo {author} {\bibfnamefont {R.}~\bibnamefont {Messerly}}, \bibinfo {author} {\bibfnamefont {Y.~W.}\ \bibnamefont {Li}}, \bibinfo {author} {\bibfnamefont {A.~I.}\ \bibnamefont {Boldyrev}}, \bibinfo {author} {\bibfnamefont {K.}~\bibnamefont {Barros}}, \bibinfo {author} {\bibfnamefont {O.}~\bibnamefont {Isayev}},\ and\ \bibinfo {author} {\bibfnamefont {S.}~\bibnamefont {Tretiak}},\ }\bibfield  {title} {\bibinfo {title} {Extending machine learning beyond interatomic potentials for predicting molecular properties},\ }\href {https://doi.org/10.1038/s41570-022-00416-3} {\bibfield  {journal} {\bibinfo  {journal}
  {Nature Reviews Chemistry}\ }\textbf {\bibinfo {volume} {6}},\ \bibinfo {pages} {653} (\bibinfo {year} {2022})}\BibitemShut {NoStop}%
\bibitem [{\citenamefont {Prodan}\ and\ \citenamefont {Kohn}(2005)}]{Prodan2005}%
  \BibitemOpen
  \bibfield  {author} {\bibinfo {author} {\bibfnamefont {E.}~\bibnamefont {Prodan}}\ and\ \bibinfo {author} {\bibfnamefont {W.}~\bibnamefont {Kohn}},\ }\bibfield  {title} {\bibinfo {title} {Nearsightedness of electronic matter},\ }\href {https://doi.org/10.1073/pnas.0505436102} {\bibfield  {journal} {\bibinfo  {journal} {Proceedings of the National Academy of Sciences}\ }\textbf {\bibinfo {volume} {102}},\ \bibinfo {pages} {11635} (\bibinfo {year} {2005})},\ \bibinfo {note} {publisher: Proceedings of the National Academy of Sciences}\BibitemShut {NoStop}%
\bibitem [{\citenamefont {DiStasio}\ \emph {et~al.}(2012)\citenamefont {DiStasio}, \citenamefont {Von~Lilienfeld},\ and\ \citenamefont {Tkatchenko}}]{DiStasio2012}%
  \BibitemOpen
  \bibfield  {author} {\bibinfo {author} {\bibfnamefont {R.~A.}\ \bibnamefont {DiStasio}}, \bibinfo {author} {\bibfnamefont {O.~A.}\ \bibnamefont {Von~Lilienfeld}},\ and\ \bibinfo {author} {\bibfnamefont {A.}~\bibnamefont {Tkatchenko}},\ }\bibfield  {title} {\bibinfo {title} {Collective many-body van der {Waals} interactions in molecular systems},\ }\href {https://doi.org/10.1073/pnas.1208121109} {\bibfield  {journal} {\bibinfo  {journal} {Proceedings of the National Academy of Sciences}\ }\textbf {\bibinfo {volume} {109}},\ \bibinfo {pages} {14791} (\bibinfo {year} {2012})}\BibitemShut {NoStop}%
\bibitem [{\citenamefont {Welborn}\ \emph {et~al.}(2018)\citenamefont {Welborn}, \citenamefont {Cheng},\ and\ \citenamefont {Miller}}]{Welborn2018}%
  \BibitemOpen
  \bibfield  {author} {\bibinfo {author} {\bibfnamefont {M.}~\bibnamefont {Welborn}}, \bibinfo {author} {\bibfnamefont {L.}~\bibnamefont {Cheng}},\ and\ \bibinfo {author} {\bibfnamefont {T.~F.}\ \bibnamefont {Miller}},\ }\bibfield  {title} {\bibinfo {title} {Transferability in machine learning for electronic structure via the molecular orbital basis},\ }\href {https://doi.org/10.1021/acs.jctc.8b00636} {\bibfield  {journal} {\bibinfo  {journal} {Journal of Chemical Theory and Computation}\ }\textbf {\bibinfo {volume} {14}},\ \bibinfo {pages} {4772} (\bibinfo {year} {2018})},\ \bibinfo {note} {publisher: American Chemical Society (ACS)}\BibitemShut {NoStop}%
\bibitem [{\citenamefont {Zulueta}\ \emph {et~al.}(2022)\citenamefont {Zulueta}, \citenamefont {Tulyani}, \citenamefont {Westmoreland}, \citenamefont {Frisch}, \citenamefont {Petersson}, \citenamefont {Petersson},\ and\ \citenamefont {Keith}}]{Zulueta2022}%
  \BibitemOpen
  \bibfield  {author} {\bibinfo {author} {\bibfnamefont {B.}~\bibnamefont {Zulueta}}, \bibinfo {author} {\bibfnamefont {S.~V.}\ \bibnamefont {Tulyani}}, \bibinfo {author} {\bibfnamefont {P.~R.}\ \bibnamefont {Westmoreland}}, \bibinfo {author} {\bibfnamefont {M.~J.}\ \bibnamefont {Frisch}}, \bibinfo {author} {\bibfnamefont {E.~J.}\ \bibnamefont {Petersson}}, \bibinfo {author} {\bibfnamefont {G.~A.}\ \bibnamefont {Petersson}},\ and\ \bibinfo {author} {\bibfnamefont {J.~A.}\ \bibnamefont {Keith}},\ }\bibfield  {title} {\bibinfo {title} {A {Bond}-{Energy}/{Bond}-{Order} and {Populations} {Relationship}},\ }\href {https://doi.org/10.1021/acs.jctc.2c00334} {\bibfield  {journal} {\bibinfo  {journal} {Journal of Chemical Theory and Computation}\ }\textbf {\bibinfo {volume} {18}},\ \bibinfo {pages} {4774} (\bibinfo {year} {2022})}\BibitemShut {NoStop}%
\bibitem [{\citenamefont {Huang}\ \emph {et~al.}(2023)\citenamefont {Huang}, \citenamefont {von Rudorff},\ and\ \citenamefont {von Lilienfeld}}]{Huang2023}%
  \BibitemOpen
  \bibfield  {author} {\bibinfo {author} {\bibfnamefont {B.}~\bibnamefont {Huang}}, \bibinfo {author} {\bibfnamefont {G.~F.}\ \bibnamefont {von Rudorff}},\ and\ \bibinfo {author} {\bibfnamefont {O.~A.}\ \bibnamefont {von Lilienfeld}},\ }\bibfield  {title} {\bibinfo {title} {The central role of density functional theory in the {AI} age},\ }\href {https://doi.org/10.1126/science.abn3445} {\bibfield  {journal} {\bibinfo  {journal} {Science}\ }\textbf {\bibinfo {volume} {381}},\ \bibinfo {pages} {170} (\bibinfo {year} {2023})},\ \bibinfo {note} {publisher: American Association for the Advancement of Science (AAAS)}\BibitemShut {NoStop}%
\bibitem [{\citenamefont {Stone}(1982)}]{Stone1982}%
  \BibitemOpen
  \bibfield  {author} {\bibinfo {author} {\bibfnamefont {C.~J.}\ \bibnamefont {Stone}},\ }\bibfield  {title} {\bibinfo {title} {Optimal {Global} {Rates} of {Convergence} for {Nonparametric} {Regression}},\ }\href {https://www.jstor.org/stable/2240707} {\bibfield  {journal} {\bibinfo  {journal} {The Annals of Statistics}\ }\textbf {\bibinfo {volume} {10}},\ \bibinfo {pages} {1040} (\bibinfo {year} {1982})},\ \bibinfo {note} {publisher: Institute of Mathematical Statistics}\BibitemShut {NoStop}%
\bibitem [{\citenamefont {Banjafar}\ and\ \citenamefont {von Rudorff}(2025{\natexlab{b}})}]{Banjafar2025a}%
  \BibitemOpen
  \bibfield  {author} {\bibinfo {author} {\bibfnamefont {A.}~\bibnamefont {Banjafar}}\ and\ \bibinfo {author} {\bibfnamefont {G.~F.}\ \bibnamefont {von Rudorff}},\ }\href {https://doi.org/10.5281/ZENODO.15796395} {\bibinfo {title} {{NablaChem}/nablachem: v25.1}} (\bibinfo {year} {2025}{\natexlab{b}})\BibitemShut {NoStop}%
\end{thebibliography}%
\end{document}